\documentclass[graybox,floatfix]{svmult}

\usepackage{helvet}         
\usepackage{courier}        
\usepackage{type1cm}        
\usepackage{makeidx}         
\usepackage{graphicx}        
\usepackage{multicol}        
\usepackage[bottom]{footmisc}

\makeindex             

\usepackage{amsmath,amssymb}
\usepackage{subfigure}
\usepackage{graphicx}
\usepackage{tabularx}
\usepackage{graphicx}
\usepackage{bm}
\usepackage{exscale}
\usepackage{amssymb}
\usepackage{amsfonts}
\usepackage{amsmath}
\usepackage{amsbsy}
\usepackage{psfrag}

\def\LSCO{La$_{2-x}$Sr$_x$CuO$_4$}
\def\LBCO{La$_{2-x}$Ba$_x$CuO$_4$}
\def\YBCO{YBa$_2$Cu$_3$O$_{6+x}$}

\def\C60{A$_x$C$_{60}$}
\def\LNSCO{La$_{1.6-x}$Nd$_{0.4}$Sr$_x$CuO$_{4}$}

\def\hty{high temperature superconductivity}
\def\hts{high temperature superconductors}

\def\SROtwo{ Sr$_{3}$Ru$_{2}$O$_{7}$}

\def\LCO{La$_2$CuO$_4$}

\def\BSCCO{Bi$_2$Sr$_2$CaCu$_2$O$_{8+\delta}$}
\def\oxychloride{Ca$_{2-x}$Na$_x$CuO$_2$Cl$_2$}
\def\LNSCO{La$_{1.6-x}$Nd$_{0.4}$Sr$_x$CuO$_{4}$}

\def\HgCu3{HgCa$_2$Cu$_3$O$_{8+y}$}
\def\HgCu4{HgBa$_2$Ca$_3$Cu$_4$O$_{10+y}$}
\def\TlCu{Tl$_2$Ba$_2$CuO$_{6+\delta}$}
\def\TlCu3{Tl$_2$Ba$_2$Ca$_2$Cu$_3$O$_{10+y}$}
\def\TlCu4{Tl$_2$Ba$_2$Ca$_3$Cu$_4$O$_{12+y}$}

\def\BiCu3{Bi$_2$Sr$_2$Ca$_{2}$Cu$_3$O$_y$}

\def\8LSCO{La$_{1.88}$Sr$_{.12}$CuO$_4$}
\def\110LNSCO{La$_{1.5}$Nd$_{0.4}$Sr$_{0.1}$CuO$_{4}$}
\def\stage4LCO{La$_{2}$CuO$_{4+\delta}$}
\def\Y248{YBa$_2$Cu$_4$O$_8$}

\def\htr{high temperature superconductor}

\def\NbSe2{NbSe$_2$}
\def\TaSe2{TaSe$_2$}
\def\TiSe2{TiSe$_2$}
\def\NaCoOH2O{Na$_{0.3}$CoO$_{2y}$H$_2$O}
\def\MgB2{MgB${}_2$}

\def\beq{\begin{equation}}
\def\eeq{\end{equation}}
\def\bea{\begin{eqnarray}}
\def\eea{\end{eqnarray}}

\renewcommand{\thefootnote}{\fnsymbol{footnote}}

\begin{document}

\title*{Electronic Liquid Crystal Phases in Strongly Correlated Systems}
\author{Eduardo Fradkin}
\institute{
Department of Physics, University of Illinois at Urbana-Champaign, 1110 West Green Street, Urbana, Illinois 61801-3080, USA, \email{efradkin@illinois.edu}}
\maketitle
\abstract{
I discuss the electronic liquid crystal (ELC) phases in correlated electronic systems, what these phases are
 and in what context they arise. I will go over the
strongest experimental evidence for these phases in a variety of systems: the two-dimensional electron
gas (2DEG) in magnetic fields, the bilayer material {\SROtwo} (also in magnetic fields), and a set of
phenomena in the cuprate superconductors (and more recently in the pnictide materials) that can be most
simply understood in terms of ELC phases. Finally we will go over the theory of these phases, focusing
on effective field theory descriptions and some of the known mechanisms that may give rise to these
phases in specific models. 
}
\date{\today}


\titlerunning{Electronic Liquid Crystal Phases}

\renewcommand{\thefootnote}{\arabic{footnote}}
\setcounter{footnote}{0}
\section[Electronic Liquid Crystal Phases]{Electronic Liquid Crystal Phases}
\label{sec:phase}

Electronic liquid crystal phases\cite{kivelson-1998} are states of correlated quantum electronic systems 
that break spontaneously either
rotational invariance or translation invariance. Since most correlated electronic systems arise in a solid state
environment the underlying crystal symmetry plays a role as it is the unbroken symmetry of the system. Thus in
practice these phases break the point group symmetry of the underlying lattice, 
in addition of the possible breaking of
the lattice translation symmetry. 

This point of view is commonplace in the classification of phases of {\em classical} liquid
crystals.\cite{degennes-1993} Classical liquid crystal systems are assemblies of a macroscopically large number of
molecules with various shapes. The shapes of the individual molecules (the ``nematogens'') affects their mutual
interactions, as well as enhancing entropically-driven interactions (``steric forces'') which, 
when combined, give rise
to the dazzling phase diagrams of liquid crystals and the fascinating properties of their phases.\cite{degennes-1993}

The physics of liquid crystals is normally regarded as part of ``soft'' condensed matter physics, while the physics of
correlated electrons is usually classified as part of ``hard'' condensed matter physics. The necessity to use both
points of view clearly brings to the fore the underlying unity of Physics as a science. Thus, one may think of this
area as ``soft quantum matter'' or quantum soft matter'' depending on to which tribe you belong to.

These lectures are organized as follows. In section \ref{sec:phase} ELC phases and their symmetries are described. 
In section \ref{sec:evidence}  I cover the main experimental evidence for these phases in 2DEGs, in {\SROtwo} and in {\hts}. 
In section \ref{sec:theories} I present the theories of stripe phases, in section \ref{sec:optimal} the relation between electronic inhomogeneity and {\hty} is 
discussed, and section{\ref{sec:pdw} is devoted to the theory of the pair density wave (the striped superconductor). Section \ref{sec:nematic} is devoted to 
the theories of nematic phases and a theory of nematic electronic order in the strong coupling regime is discussed in section \ref{sec:strong-coupling}.   
The stripe-nematic quantum phase transition is discussed in section \ref{sec:NSQCP}.

\subsection{Symmetries of Electronic Liquid Crystal Phases}
\label{sec:symmetries}

\begin{figure}[hbt]
\begin{center}
\includegraphics[width=0.5\textwidth]{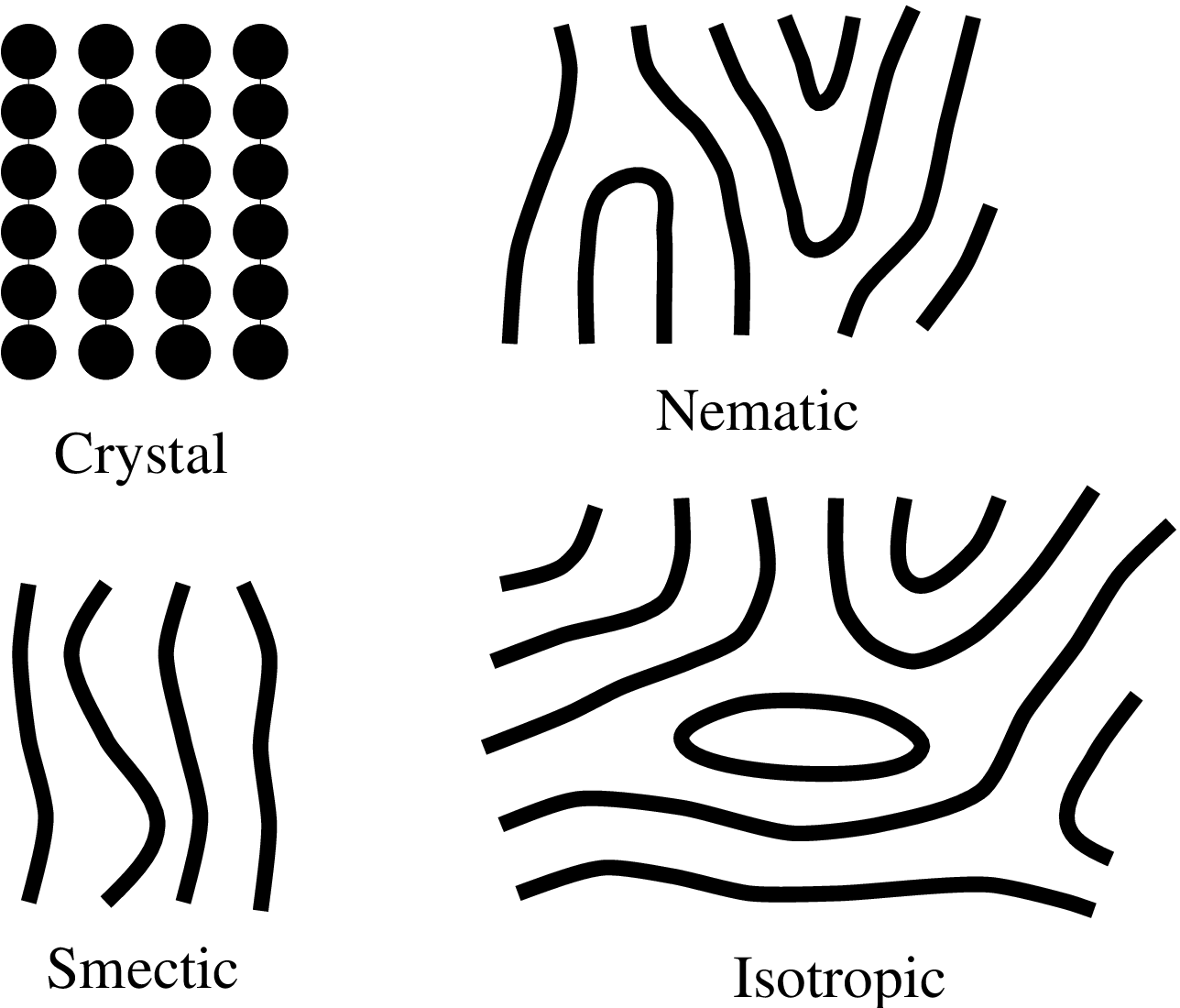}
\end{center}
\caption{Cartoon of liquid crystal phases. }
\label{fig:elc}
\end{figure}

We will follow Ref.\cite{kivelson-1998} and classify the ELC phases of strongly 
correlated electrons\footnote{You may call the ELC phases the anisotropic states of point
particles!} following
the symmetry-based scheme used in classical liquid crystals\cite{degennes-1993,chaikin-1995}:
\begin{enumerate}
\item
{\em Crystalline phases}: phases that break all continuous translation symmetries and rotational invariance.
\item
{\em Smectic (``stripe'') phases}: phases break one translation symmetry and rotational invariance.
\item
{\em Nematic and hexatic phases}: uniform (liquid) phases that break rotational invariance.
\item
{\em Isotropic}: uniform and isotropic phases.
\end{enumerate}
A cartoon of the real space structure of
these ordered phases is shown in Fig.\ref{fig:elc}.

Unlike classical liquid crystals, electronic systems carry charge and spin, and have strong quantum mechanical effects
(particularly in the strong correlation regime). 
This leads to a host of interesting possibilities of ordered states in which the liquid crystalline character of the 
spatial structure of these states becomes intertwined with the ``internal'' degrees of freedom of electronic systems. 
These novel ordered
phase will be the focus of these lectures. One of the aspects we will explore is the structure of their phase diagrams. 
Thus in addition
of considering the thermal melting of these phases, we will also be interested in the 
{\em quantum} melting of these states and the
associated quantum phase transitions. (see a sketch in Fig.\ref{fig:elc-phase-diagram}.)

\begin{figure}[t!]
\psfrag{T}{$T$}
\psfrag{r}{$r$}
\psfrag{Crystal}{Crystal}
\psfrag{Nematic}{Nematic}
\psfrag{Smectic}{Smectic}
\psfrag{Superconductor}{Superconductor}
\begin{center}
\includegraphics[width=0.6\textwidth]{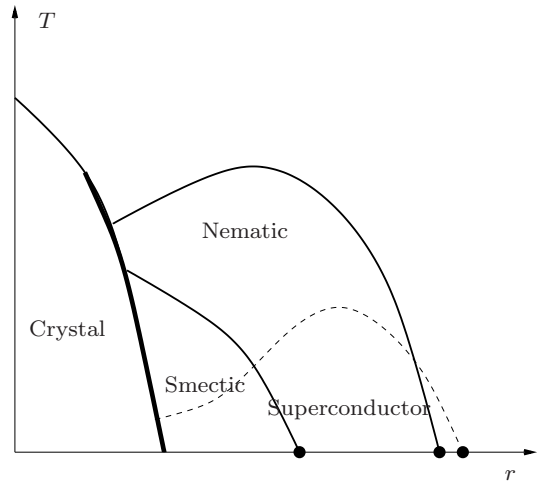}
\end{center}
\caption{Schematic phase diagram of electronic liquid crystal phases. Here $T$ is temperature and $r$ denotes a tuning parameter 
the controls the strength
of the quantum fluctuations. In practice it can represent doping, magnetic field, pressure or even material.
The full dots are quantum critical points.}
\label{fig:elc-phase-diagram}
\end{figure}

In this context, the {\em crystalline phases} are either insulating or ``almost
insulating'', {\it e.g.\/} multiple charge density density waves (CDW) ordered states either commensurate or 
sliding (incommensurate).
However, these phase may also be superconducting either by coexistence or, more interestingly, by modulating 
the superconducting states
themselves.
Similarly, electron nematics are anisotropic metallic or superconducting states, while the isotropic phase 
are also either metallic or
superconducting. As we will see these phases display a set of rather striking and unusual behaviors, some 
of which have been observed in
recent experiments.

\subsection{Order Parameters and their Symmetries}
\label{sec:OP-symmetries}

The order parameters of ELC phases are well known.\cite{kivelson-1998,kivelson-2003} In the crystalline phases, the order parameters are 
$\rho_{\vec{K}}$, 
the expectation values of the
density operators at the set of ordering wave vectors $\{ \vec{K}\}$ the defines the crystal.\cite{chaikin-1995}
\begin{equation}
 \rho_{\vec{K}}=\int d \vec{r} \; \rho(\vec{r}) \; e^{i {\vec{K} \cdot \vec{r}}}
 \label{eq:rhoK}
 \end{equation}
 where $\rho(\vec{r})$ is the local charge density. Thus, under an uniform translation by $\vec{R}$, $\rho_{\vec{K}}$
 transforms as
 \begin{equation}
 \rho_{\vec{K}} \to \rho_{\vec{K}} \; e^{i {\vec{K}} \cdot {\vec{R}}}
 \label{eq:rhok-transformation}
 \end{equation}
Smectic phases are unidirectional density waves and
their order parameters are also expectation values $\rho_{\bf K}$ but for {\em only one} wave vector $\vec{K}$. For charged
systems, $\rho(\vec{r})$ is the {\em charge density}, and the order parameter $\rho_{\vec{K}}$ is the {\em charge density
wave} (CDW) order parameter. Since $\rho(\vec{r})$ is {\em real}, $\rho_{\vec{K}}=\rho_{-{\vec{K}}}^*$, and the density can be
expanded as
\begin{equation}
\rho(\vec{r})=\rho_0(\vec{r})+\rho_{\vec{K}}(\vec{r}) e^{i {\vec{K}}\cdot {\vec{r}}}+ {\rm c.c.}
\label{eq: density-expansion}
\end{equation}
where $\rho_0(\vec{r})$ are the Fourier components {\em close} to zero wave vector, $\vec{k}=0$, and $\rho_{\vec{K}}(\vec{r})$
are the Fourier components with wave vectors {\em close} to $\vec{k}=\vec{K}$. Hence, a density wave (a {\em smectic})
 is represented by an
{\em complex} order parameter field, in this case $\rho_{\vec{K}}(\vec{r})$. This is how we will describe a CDW and a charge
stripe (which from the point of view of symmetry breaking have the same description).\footnote{On the other hand, 
in the case of a crystal phase,
the expansion is 
\begin{equation}
\rho(\vec{r})=\rho_0(\vec{r})+\sum_{{\vec{K}} \in \Gamma} \rho_{\vec{K}}(\vec{r}) e^{i {\vec{K}}\cdot {\vec{r}}}+ 
{\rm c.c.}
\end{equation}
where $\Gamma$ denotes the set of primitive lattice vectors of the crystal phase.\cite{chaikin-1995}} 

Smectic order is detected most easily in scattering experiments through the measurement of the {\em static structure factor}, usually denoted by 
$S(\vec{k})$,
\begin{equation}
S(\vec{k})=\int \frac{d\omega}{2\pi} S(\vec{k},\omega)
\label{eq:S(k)}
\end{equation}
where $S(\vec{k},\omega)$ is the {\em dynamical structure factor}, {\it i.e.\/} the Fourier transform of the (in this case) density-density correlation function. 
The signature of smectic order is the existence of delta-function component of $S(\vec{k})$ at the ordering wave vector, $\vec{k}=\vec{K}$, with a prefactor 
that is equal to $|\langle \rho_{\vec{K}}\rangle|^2$.\cite{chaikin-1995}

In the case of a {\em spin density wave} (a
``spin stripe'') the picture is the same except that the order parameter field is multi-component, $\vec{S}_{\vec{K}}(\vec{r})$, corresponding to different spin 
polarizations. Thus, the local spin density ${\bf S}(\bf r)$ has the expansion
\begin{equation}
\vec{S}(\vec{r})=\vec{S}_0(\vec{r})+ \vec{S}_{\vec{K}}(\vec{r}) \; e^{i {\vec{K}} \cdot {\vec{r}}}+ {\rm c.c.}
\label{eq:sdw}
\end{equation}
where $\vec{S}_0(\vec{r})$ denotes the local (real) {\em ferromagnetic} order parameter and $\vec{S}_{\vec{K}}(\vec{r})$ is the
(complex) SDW (or
spin stripe) order parameter field, a complex vector in spin space.

One of the questions we will want to address is the connection between these orders and superconductivity. The
superconducting order parameter, a pair condensate, is the {\em complex} field $\Delta(\vec{r})$. It is natural
(and as we will see borne out by current experiments) to consider the case in which the superconducting order is also
modulated, and admits an expansion of the form
\begin{equation}
\Delta(\vec{r})=  {\Delta}_0 (\vec{r})+ {\Delta}_{\vec{K}}(\vec{r})\;   e^{i {\vec{K}} \cdot {\vec{r}}}+ 
{\Delta}_{ -{\vec{K}}} (\vec{r})\;   e^{-i {\vec{K}} \cdot {\vec{r}}}
\label{eq:pdw}
\end{equation}
where the uniform component $\Delta_0$ is the familiar BCS order parameter, and $\Delta_{\bf K}(\bf r)$ is the {\em
pair-density-wave} (PDW) order parameter,\cite{berg-2007,berg-2008} closely related 
to the Fulde-Ferrell-Larkin-Ovchinnikov (FFLO) 
order parameter\cite{Fulde-1964,Larkin-1964} (but without an external magnetic field). Since $\Delta(\bf r)$ is complex, 
$\Delta_{\vec{K}}(\vec{r}) \neq \Delta_{-\vec{K}} (\vec{r})^*$, the PDW state has two complex order parameters.\footnote{I will
not discuss the case of spiral order here.}

In contrast, 
nematic phases are translationally invariant but break rotational invariance. Their order parameters 
 transform irreducibly under the rotation group for a continuous system, or under the point (or space) group of the
 lattice. Hence, the order parameters of a
nematic phase (hexatic and their generalizations) are symmetric traceless tensors, that we will denote by $Q_{ij}$.
\cite{degennes-1993} In 2D
as most of the problems we will be interested in are 2D systems (or quasi2D systems), 
the order parameter takes the form (with $i,j=x,y$)
\begin{equation}
Q_{ij}=
\begin{pmatrix}
Q_{xx} & Q_{xy}\\
Q_{xy} & -Q_{xx}
\end{pmatrix}
\label{eq:Q2D}
\end{equation}
 which, alternatively, can be written in terms of a {\em director} $N$,
 \begin{equation}
 N=Q_{xx}+iQ_{xy}=|N| \; e^{i \varphi}
 \label{eq:N}
 \end{equation}
 Under a rotation by a fixed angle $\theta$, 
 $N$ transforms as\footnote{For a lattice system, rotational symmetries are those of the point 
 (or space) group symmetry of the lattice. thus, nematic
order parameters typically become Ising-like (on a square lattice) or three-state Potts on a triangular lattice (and so
forth.)}
 \begin{equation}
 N \to N \; e^{i 2 \theta}
 \label{eq:N-transformation}
 \end{equation}
 Hence, it changes sign under a rotation by $\pi/2$ and it is {\em invariant} under a rotation by $\pi$ (hence the name
 {\em director}, a headless vector). On the other hand,  it is  invariant under uniform translations by $\vec{R}$.

In practice we will have a great latitude when choosing a nematic order parameter as any symmetric traceless tensor in
space coordinates will transform properly. In the case of a charged metallic system, a natural choice to describe a
metallic nematic state is the traceless symmetric component of the {\em resistivity} (or conductivity)
tensor.\cite{fradkin-2000,cooper-2002,ando-2002,borzi-2007} In 2D we
will use
\begin{equation}
Q_{ij}=
\begin{pmatrix}
\rho_{xx}-\rho_{yy} & \rho_{xy}\\
\rho_{xy} & \rho_{yy}-\rho_{xx}
\end{pmatrix}
\label{eq:N-resistivity}
\end{equation}
where $\rho_{xx}$ and $\rho_{yy}$ are the {\em longitudinal} resistivities and $\rho_{xy}=\rho_{yx}$ is the transverse
({\em Hall}) resistivity. A similar analysis can be done in terms of the dielectric tensor, 
which is useful in the context
of light scattering experiments. 
On the other hand, when looking at the spin polarization properties of a system other measures
of nematic order are available. For instance, in a neutron
scattering experiment, the anisotropy under a rotation $\mathcal{R}$ (say, by $\pi/2$) of the the structure factor $S(\vec{k})$
\begin{equation}
Q\sim S(\vec{k})-S(\mathcal{R} \vec{k})
\label{eq:QR}
\end{equation}
is a measure of the nematic order parameter $Q$.\cite{kivelson-2003,hinkov-2007}.

Other, more complex, yet quite interesting phases are possible.  One should keep in mind that the nematic order parameter
(as defined above) corresponds to a field that transforms under the lowest (angular momentum $\ell=2$) 
irreducible representation of the 
group rotations, compatible with inversion symmetry. The nematic phase thus defined has $d$-wave symmetry, the symmetry of
a quadrupole. Higher symmetries are also possible, {\it e.g.\/} hexatic ($\ell=6$). However it is also possible to have
states that break both rotational invariance and 2D inversion (mirror reflection), as in the $\ell=3$ channel. Such states
break (although mildly) time-reversal invariance.\cite{sun-2008b,varma-2005}  Other complex phases arise 
by combining the nematic 
order in real space with those of some internal symmetry, {\it e.g.\/} spin or orbital degeneracies. 
Thus one can consider nematic 
order parameters in the {\em spin-triplet} channels, which give rise to a host of (as yet undetected) phases with
fascinating behaviors in the spin channel or under time-reversal, such as the dynamical generation of spin orbit coupling
or the spontaneous breaking of time-reversal invariance.\cite{wu-2004,wu-2007,sun-2008b}

\subsection{Electronic Liquid Crystal Phases and Strong Correlation Physics}

One of the central problems in condensed matter physics is the understanding of doped Mott insulators. 
Most of the interesting systems in
condensed matter, notably {\hts}, are doped Mott insulators.\cite{anderson-1987}
A Mott insulator is a phase of an electronic system in which there is a gap in the single 
particle spectrum due to the effects of electronic correlations and not to features of the band structure. 
Thus, Mott insulators have an odd number of electrons in the unit cell. For a system like this, band theory would
predict that such systems must be metallic, not insulating, and be described by the Landau theory of the
Fermi liquid. Electronic systems that become insulating due to the effects of strong correlation are states of matter with non-trivial
correlations. 

Most known Mott insulating states are ordered phases, associated with the spontaneous breaking of 
some global symmetry of the electronic
system, and have a clearly defined order parameter. Typically the Mott state is an 
antiferromagnetic state (or generalizations thereof).
However there has been a sustained interest possible in non-magnetic Mott phases, 
{\it e.g.\/} dimerized, various sorts of conjectured
spin liquids, etc., some of which do not admit an order parameter description 
(as in the case of the topological phases.)  

We will not concern ourselves on these questions here. What will matter to us is that 
doping this insulator by holes disrupts the
correlations that define the insulating state. Consequently doped holes are more costly 
(energetically) if they are apart than if they
are together. The net effect is that the disruption of the correlations of the 
Mott state results in an effective strong attractive 
interaction between the doped holes.  This effect was early on mistaken for aa sign 
of pairing in models of {\hts} (such as the
Hubbard and $t-J$ models). Further analysis revealed that this effective attraction 
meant instead the existence of a {\em generic} 
instability of strongly correlated systems to {\em phase separation}.\cite{emery-1990} 
This feature of strong correlation has been amply
documented in numerical simulations (see, for instance, Ref.\cite{dagotto-1994}).

Due to the inherent tendency to phase separation of Hubbard-type models (and its descendants), the {\em
insulating} nature of a Mott insulator cannot be ignored and, in particular, its inability to screen the 
longer range Coulomb interactions. Thus, quite generally, one can expect that the combined effects of 
the kinetic energy of the doped
holes and the repulsive Coulomb interactions should in effect {\em frustrate} the tendency to phase 
separation of short-ranged models of
strong correlation.\cite{emery-1993} 

The existence of strong short range attractive forces and long range repulsion is a recipe for the 
formation of phases with complex
spatial structure. As noted above, this is what happens in classical liquid crystals. 
It is also the general mechanism giving rise to
generally inhomogeneous phases in classical 
complex fluids such as ferrofluids and heteropolymers,\cite{seul-1995} as well as in 
astrophysical problems such as the crusts of
neutron stars.\cite{ravenhall-1993}

The point of view that we take in these lectures is that the behavior observed in the underdoped 
regime of {\hts} and in other strongly
correlated systems is due to the strong tendency that these systems have to form generally 
inhomogeneous and anisotropic phases,
``stripes''. In the following lectures we will go over the experimental evidence for these 
phases and for their theoretical underpinning.\footnote{Ref. \cite{fradkin-2010} is a recent, complementary, review of the phenomenology of nematic phases in strongly correlated systems.}

\begin{figure}[t!]
\psfrag{T}{$T$}
\psfrag{x}{$x$}
\psfrag{antiferromagnet}{\small antiferromagnet}
\psfrag{superconductor}{\small superconductor}
\psfrag{"1/8"}{$\frac{1}{8}$}
\psfrag{pseudogap}{\small pseudogap}
\psfrag{bad-metal}{\small bad metal}
\begin{center}
\includegraphics[width=0.6\textwidth]{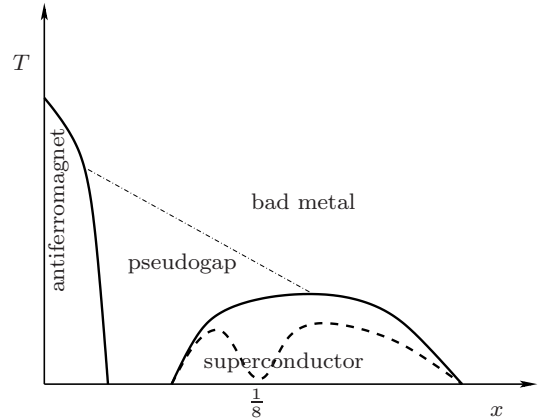}
\end{center}
\caption{Schematic phase diagram of the cuprate superconductors. 
The full lines are the phase boundaries for the antiferromagnetic and superconducting phases. 
The broken line is the phase diagram for a system with static stripe order and a pronounced $1/8$ anomaly. 
The dotted line marks the crossover between the bad metal and pseudogap regimes.}
\label{fig:htsc}
\end{figure}

\section[Experimental Evidence]{Experimental Evidence in Strongly Correlated Systems}
\label{sec:evidence}

During the past decade or so experimental evidence has been mounting of the existence of electronic 
liquid crystal phases in a variety of strongly correlated (as well as not as strongly correlated)
electronic systems. We will be particularly interested in the experiments in the copper oxide {\hts}, 
in the  ruthenate materials (notably {\SROtwo}), and in two-dimensional electron gases (2DEG) in large 
magnetic fields. However, as we will discuss below, these concepts are also relevant to more conventional 
CDW systems.

\subsection{Nematic Phases in the 2DEG in High Magnetic Fields}
\label{sec:2deg}

\begin{figure}[t!]
\begin{center}
\subfigure[]{\includegraphics[width=0.49\textwidth]{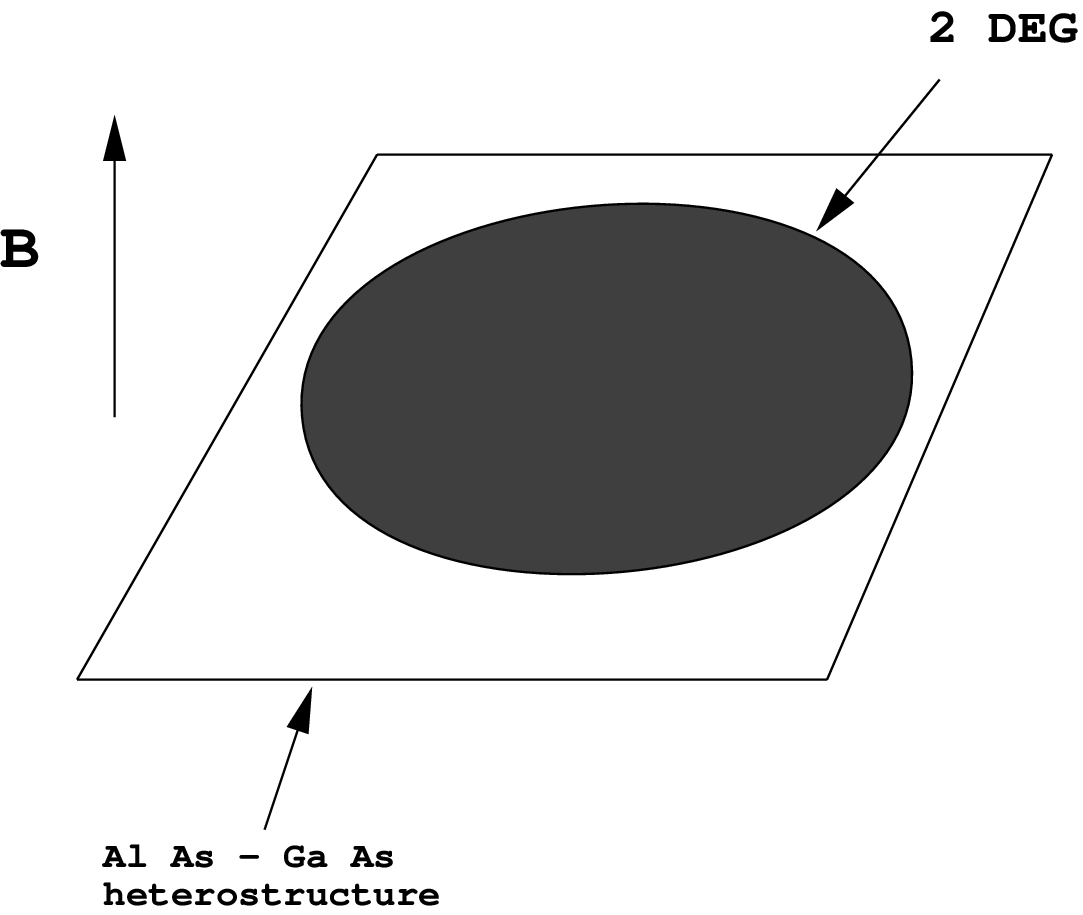}}
\subfigure[]{\includegraphics[width=0.49\textwidth]{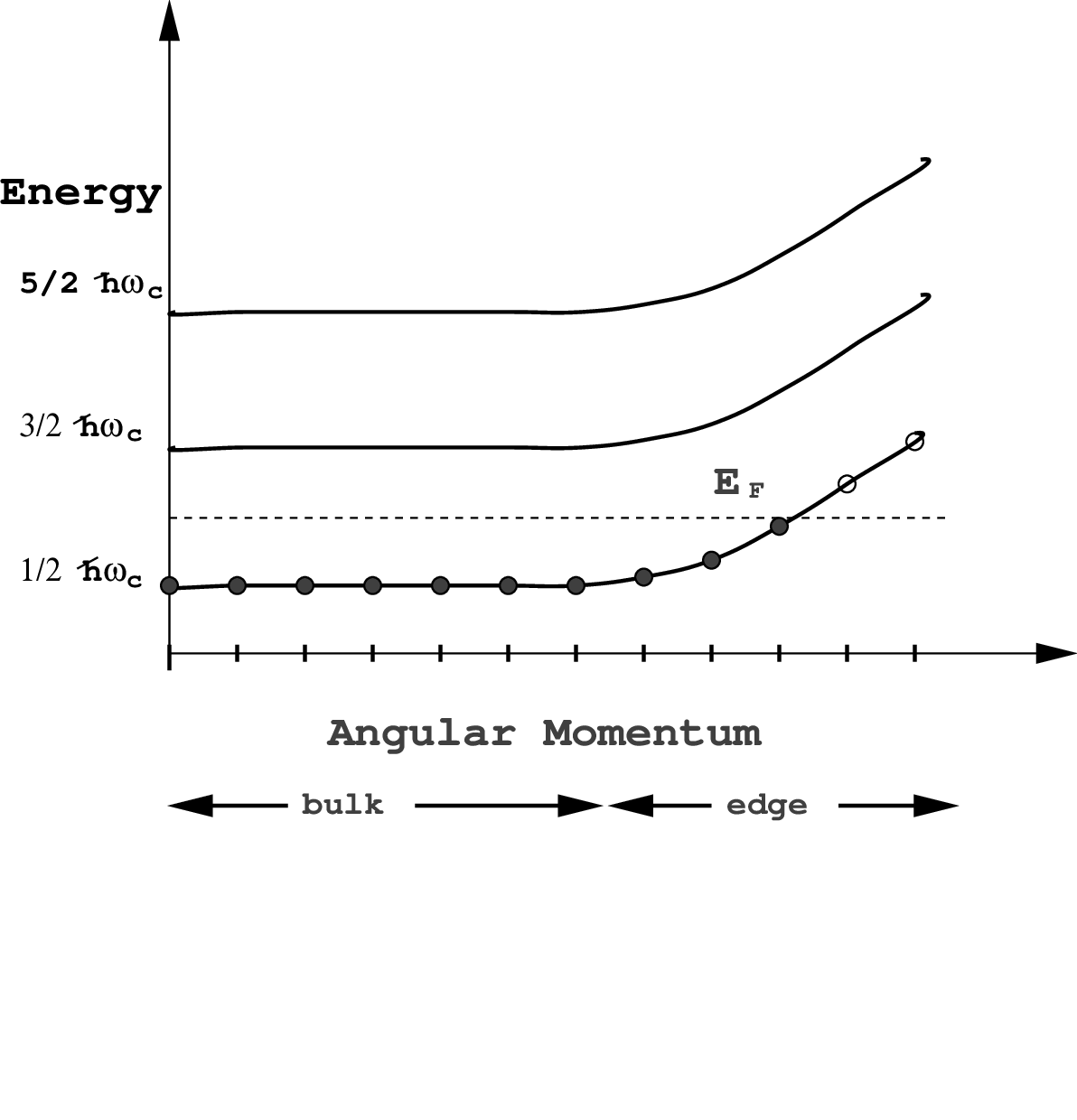}}
\end{center}
\caption{a) 2DEG in a magnetic field. b) Landau levels.}
\label{fig:qhLL}
\end{figure}

To this date, the best documented electron nematic state is the anisotropic compressible state 
observed in 2DEGs in large magnetic fields near the middle of a Landau level, with Landau index 
$N \geq 2$\cite{lilly-1999a,lilly-1999b,du-1999,pan-1999}.
In ultra high mobility samples of a 2DEG in AlAs-GaAs heterostructures,  
transport experiments in the second Landau level (and above) near the center 
of the Landau level  show a pronounced anisotropy of the longitudinal 
resistance rising sharply below $T \simeq 80$ mK, with an anisotropy 
that increases by orders of magnitude as the temperature is lowered.
This effect is only seen in ultra-clean samples, with nominal mean free paths of about 0.5 mm (!) and nominal mobilities of $10 - 30 \times 10^6$.\footnote{The anisotropy is strongly suppressed by disorder}


A nematic order parameter can be constructed
phenomenologically from the measured resistivity tensor, by taking the symmetric traceless piece of it. 
This was done in
Ref.\cite{fradkin-2000} where a fit of the data of Lilly et al\cite{lilly-1999a,lilly-1999b} was shown 
to be consistent with a
classical 2D XY model (in a weak symmetry breaking field). 
\begin{figure}[h!]
\begin{center}
\includegraphics[width=0.6\textwidth]{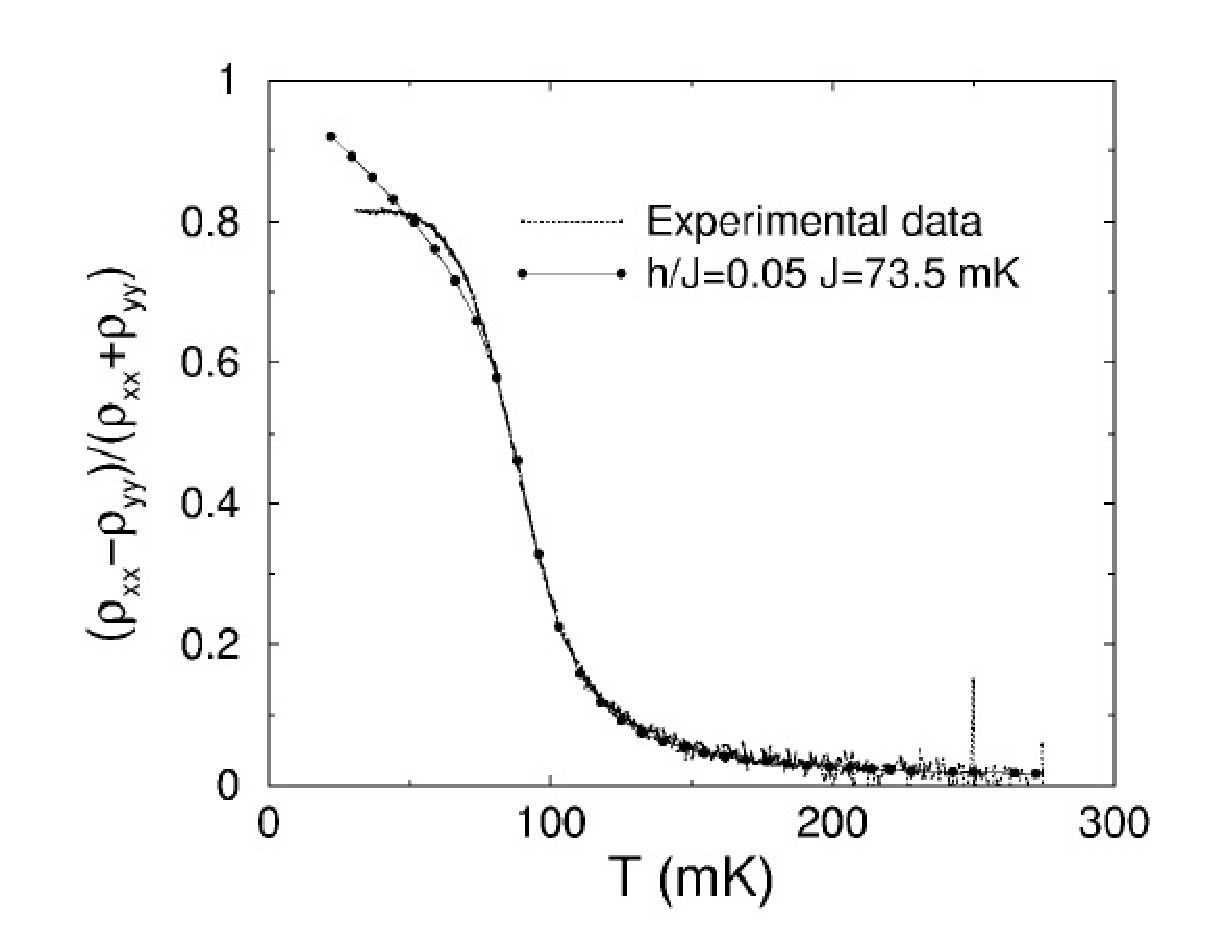}
\end{center}
\caption{
Nematic order in the 2DEG; fit of the resistance anisotropy to a 2D XY model 
Monte carlo simulation.\cite{fradkin-2000}. }
\label{fig:2dXY}
\end{figure}
A 2D XY symmetry is expected for a planar nematic order
provided the weak lattice symmetry breaking is ignored. Presumably lattice anisotropy is responsible for the
saturation shown at low temperatures in Fig.\ref{fig:2dXY}.

These experiments were originally interpreted as evidence for a quantum Hall smectic (stripe) 
phase\cite{koulakov-1996,moessner-1996,fradkin-1999,macdonald-2000,barci-2002a}. 
However, further experiments (\cite{cooper-2001,cooper-2002,cooper-2003} 
did not show any evidence of pinning of this putative unidirectional CDW as the $I-V$ curves were found to be 
strictly linear at low bias. In addition broadband noise in the current, a characteristic of CDW systems. 
In contrast, extremely sharp threshold electric fields and broadband noise in transport 
was observed in a nearby reentrant integer quantum Hall phase, suggesting a crystallized electronic state. 
These facts, together with a detailed analysis of the experimental data, suggested that the compressible 
state is in an electron nematic phase\cite{fradkin-1999,fradkin-2000,wexler-2001,radzihovsky-2002,doan-2007}, 
which is better understood as a quantum melted stripe phase.\footnote{The 2DEG in a strong magnetic field is inherently a strongly correlated system as the interaction is always much bigger than the (vanishing) kinetic energy.} An alternative picture, a nematic phase accessed by a
Pomeranchuk instability from a ``composite fermion'' Fermi liquid is conceivable but hard to justify
microscopically.\cite{oganesyan-2001,doan-2007}

\subsection{The Nematic Phase of {\SROtwo}}
\label{sec:ruthenate}


Recent magneto-transport experiments in the quasi-two-dimensional bilayer ruthenate {\SROtwo} 
by the St. Andrews group\cite{borzi-2007} have given strong evidence of a strong temperature-dependent 
in-plane transport anisotropy in strongly correlated materials at low temperatures  $T\lesssim 800$ mK and for a window 
of perpendicular magnetic fields around $7.5$ Tesla.
 {\SROtwo} is a quasi-2D bilayer material known to have a
metamagnetic transition as a function of applied perpendicular magnetic field and temperature.Contrary to the
case of the 2DEG in AlAs-GaAs heterostructures and quantum wells, the magnetic fields applied to {\SROtwo} are too weak to
produce Landau quantization. However, as in the case of the 2DEG of the previous section, the transport anisotropy appears at very
low temperatures and only in the cleanest samples. The observed transport anisotropy has a strong temperature and field
dependence (although not as pronounced as in the case of the 2DEG).
These experiments provide strong evidence that the 
system is in an electronic nematic phase in that range of magnetic fields\cite{borzi-2007,fradkin-2007}. 
The electronic nematic phase appears to have preempted a metamagnetic QCP in the same range of magnetic 
fields\cite{grigera-2001,millis-2002,perry-2004,green-2005}. 
This suggests that proximity to phase-separation may be a possible microscopic mechanism to trigger such 
quantum phase transitions, consistent with recent ideas on the role of Coulomb-frustrated phase separation 
in 2DEGs\cite{jamei-2005,lorenzana-2002b}.


\subsection{Stripe Phases and Nematic Phases in the Cuprates}
\label{sec:cuprates}

In addition to {\hty}, the copper oxides materials display a strong tendency to have charge-ordered states, 
such as  stripes. The relation between charge ordered states\cite{kivelson-2007}, as well as other proposed 
ordered states\cite{chakravarty-2001c,varma-2005}, and the mechanism(s) of {\hty} is a subject of intense 
current research. It is not, however, the main focus of these lectures. 


Stripe phases have been extensively investigated in {\hts} 
and detailed and recent reviews are available on this subject\cite{kivelson-2003,tranquada-2007}. 
Stripe phases in {\hts} have unidirectional order in both spin and charge (although not always)  
which are typically incommensurate. In general the detected stripe order 
(by low energy inelastic neutron scattering) in {\LSCO}, {\LBCO} and {\YBCO} 
(see Refs.\cite{kivelson-2003} and \cite{tranquada-2007} and references therein) is not static but ``fluctuating''. 
As emphasized in Ref.\cite{kivelson-2003}, ``fluctuating order'' means that there is no true long range unidirectional 
order. Instead, the system is in a (quantum) disordered phase, very close to a quantum phase transition to such an 
ordered phase, with very low energy fluctuations that reveal the character of the proximate ordered state. 
On the other hand, in {\LBCO} near $x=1/8$ (and in {\LNSCO} also near $x=1/8$ where they were discovered first\cite{tranquada-1995}),  the order 
detected  
by elastic neutron scattering\cite{tranquada-2004}, and resonant X-ray scattering in {\LBCO} \cite{abbamonte-2005} 
also near $x=1/8$, becomes true long range static order. 

In the case of {\LSCO}, away from $x=1/8$, and particularly on the more underdoped side,  
the in-plane resistivity has a considerable temperature-dependent anisotropy, 
which has been interpreted as an indication of electronic nematic order.\cite{ando-2002} 
The same series of experiments also showed that very underdoped {\YBCO} is an electron nematic as well. 



The most striking evidence for electronic nematic order in {\hts} are the recent  neutron scattering experiments in 
{\YBCO} at $y=6.45$\cite{hinkov-2007b}. 
In particular, the temperature-dependent anisotropy of the inelastic
 neutron scattering in {\YBCO} shows that there is a critical temperature for nematic order (with $T_c \sim 150 K$) 
 where the inelastic neutron peaks also become incommensurate. 
 Similar effects were reported by the same group\cite{hinkov-2006} at higher doping levels ($y\sim 6.6$) 
 who observed that the nematic signal was decreasing in strength suggesting the existence of a 
 nematic-isotropic quantum phase transition closer to optimal doping. 
 Fluctuating stripe order in underdoped {\YBCO} has been detected earlier on 
 in inelastic neutron scattering experiments \cite{mook-2000,stock-2004} which, 
 in hindsight, can be reinterpreted as evidence for nematic order.   
 However, as doping increases past a $y \sim 6.6$ a spin gap appears and magnetic scattering is strongly suppressed at low energies (in the absence of 
 magnetic fields) making inelastic neutron scattering experiments less effective in this regime.

 In a particularly series of interesting experiments, the Nernst coefficient was measured in {\YBCO} ranging from the very underdoped regime, where 
 inelastic neutron scattering detects nematic order, to a slightly overdoped regime.\cite{taillefer-nematic-2009}
 The Nernst coefficient is defined as follows. Let $j_e$ and $j_Q$ be the charge and heat currents established in a 2D sample by an 
 electric field $\vec{E}$ 
 and a temperature gradient $\vec{\nabla} T$:
 \beq
 \begin{pmatrix}
 j_e\\
 j_Q
 \end{pmatrix}
 =
 \begin{pmatrix}
 \vec{\sigma} & \vec{\alpha}\\
T \vec{\alpha}&\vec{\kappa}
 \end{pmatrix}
 \begin{pmatrix}
 \vec{E}\\
 -\vec{\nabla} T
 \end{pmatrix}
 \label{eq:lrt}
 \eeq
 where $\vec{\sigma}$, $\vec{\alpha}$ and $\vec{\kappa}$ are $2 \times 2$ tensors for the conductivity, the thermoelectric conductivity and the 
 thermal conductivity respectively. The Nernst coefficient, also a $2 \times 2$ tensor $\vec{\theta}$ is measured (see Ref.\cite{ong-2007}) say by 
 applying a temperature gradient in the $x$ direction and measuring the voltage along the $y$ direction:
 \beq
 \vec{E}=-\vec{\theta} \vec{\nabla} T
 \label{eq:nernst}
 \eeq
 Since no current flows through the system, the Nernst tensor is
 \beq
 \vec{\theta}=- \vec{\sigma}^{-1} \vec{\alpha}
 \label{eq:nernst2}
 \eeq
 These experiments revealed that the Nernst (tensor) coefficient has an anisotropic component whose onset coincides (within the error bars) with the 
 conventionally defined value of the pseudogap temperature $T^*$, and essentially tracks its evolution as a function of doping. Thus, it appears that,  at 
 least in {\YBCO}, the pseudogap is a regime with nematic order 
 The same group had shown earlier than the Nernst 
 coefficient is a sensitive indicator of the onset of stripe charge order in {\LNSCO}\cite{cyrchoiniere-2009}.


Inelastic neutron scattering experiments have found  nematic order also in \\
{\LSCO} materials where 
fluctuating stripes where in fact first discovered \cite{tranquada-1995}. Matsuda {\it et al} \cite{matsuda-2008} 
have found  in  underdoped {\LSCO} ($x=0.05$), a material that was known to 
have ``fluctuating diagonal stripes'', evidence for nematic order similar to what Hinkov et al\cite{hinkov-2007b} found in underdoped {\YBCO}. Earlier experiments in {\LSCO} in moderate magnetic fields had also shown that a
spin stripe state became static over some critical value of the field.\cite{lake-2002}
These experiments strongly suggest that the experiments that had previously identified the {\hts} as having  
``fluctuating stripe order'' (both inside and outside  the superconducting phase) were most likely detecting an 
electronic nematic phase, quite close to a state with long range stripe (smectic) order. In all cases the background 
anisotropy (due to the orthorhombic distortion of the crystal structure) acts as a symmetry breaking field that 
couples linearly to the nematic order, thus rounding the putative thermodynamic transition to a state with spontaneously 
broken point group symmetry. These effects are much more apparent at  low doping where the crystal orthorhombicity 
is significantly weaker.


\begin{figure}[h!]
\begin{center}
\includegraphics[width=0.5\textwidth]{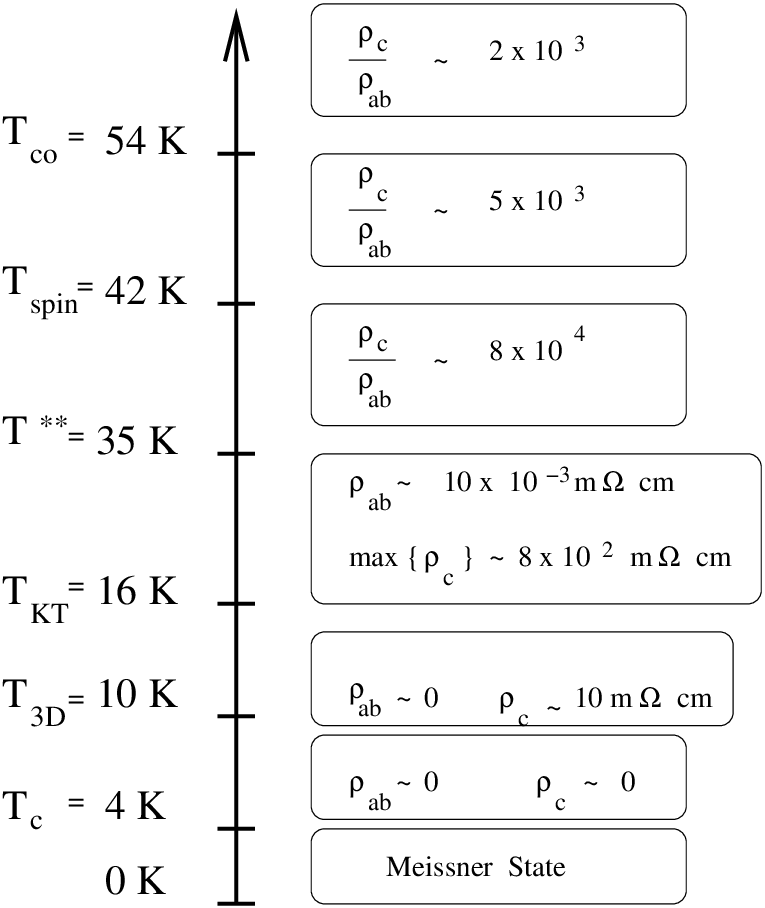}
\end{center}
\caption{Summary of the behavior of the stripe-ordered superconductor {\LBCO} near $x=1/8$: 
$T_{\rm co}$ is the charge ordering temperature, $T_{\rm spin}$ the spin ordering temperature, 
$T^{**}$ marks the beginning of layer decoupling behavior, $T_{KT}$ is the 2D superconducting 
temperature (``KT''), $T_{3D}$ is the 3D resistive transition, and  $T_c$ is the 3D Meissner 
transition.}
\label{fig:scales-lbco}
\end{figure}

In {\LBCO} at $x=1/8$ there is strong evidence for a complex stripe ordered state that intertwines 
charge, spin and superconducting order\cite{li-2007,berg-2007}. 
In fact {\LBCO} at $x=1/8$
appears to have some rather fascinating properties. As summarized in Fig.\ref{fig:scales-lbco}, {\LBCO} at $x=1/8$ has a
very low critical superconducting $T_c \sim 4$K (where the Meissner state sets in). However it is known from
angle-resolved photoemission (ARPES) experiments that the anti-nodal gap (which roughly gives the pairing scale) is
actually largest at $x=1/8$\cite{valla-2006} (or unsuppressed by the $1/8$ anomaly according to Ref.\cite{he-2008}.)
Static charge stripe order sets in at 54K (where there is a structural transition from the LTO to the LTT lattice structure), but static spin stripe order
only exists below 42 K. As soon as static spin order sets in, the in-plane resistivity begins to decrease very rapidly
with decreasing temperature, while the c-axis resistivity increases. 
Below 35 K strong
2D superconducting fluctuations are observed and at 16 K the in-plane resistivity vanishes at what appears to be a
Kosterlitz-Thouless transition. 
However, the full 3D resistive 
transition is only reached at 10 K (where $\rho_c \to 0$)
although the Meissner state is only established below 4K! This dazzling set of phenomena shows clearly that spin, charge
and superconducting order are forming a novel sort of {\em intertwined} state, rather than {\em compete}. We have
conjectured that a pair density wave (PDW) is stabilized in this intermediate temperature
regime\cite{berg-2007,berg-2008,berg-2009}. Similar phenomenology, {\it i.e.\/} a dynamical layer decoupling, has been seen
in {\LSCO} at moderate fields where the stripe order is static.\cite{schafgans-2008} 
We will return below on how a novel state, the pair-density wave, explains these phenomena.


An important caveat to the analysis  we presented here is that in doped systems there is always quenched disorder, 
which has different degrees of short range ``organization'' in different {\hts}. 
Since disorder also couples linearly to the charge order parameters it ultimately 
also rounds the transitions and render the system to a glassy state 
(as noted already in Refs.\cite{kivelson-1998,kivelson-2003}). Such effects are evident in scanning tunneling microscopy 
(STM) experiments in {\BSCCO} which revealed that the high energy (local) behavior of the {\hts} has 
charge order and it is glassy\cite{howald-2003a,kivelson-2003,hanaguri-2004,kohsaka-2007,vershinin-2004}. This is most remarkable as the STM data on {\BSCCO} at low bias shows quasiparticle propagation in the superconducting state (but not above $T_c$). Yet, at high bias ({\it i.e.\/} high energies) there are no propagating ``quasiparticles'' but, instead, provides a vivid image of electronic inhomogeneity with short range charge  order. 
This behavior is contrary to what is commonly the case in conventional superconductors where STM at high energies shows Fermi-liquid like electronic quasiparticles. Similarly, the high energy spectrum of ARPES has never resembled that of a conventional metal. 
we note that a recent analysis of this data by Lawler and coworkers has revealed the existence of nematic order over much longer length scales than the broken positional order \cite{lawler-2009b}.

Finally, we note that in the recently discovered iron pnictides based family of {\hts}, 
such as La (O$_{1-x}$F$_x$)FeAs and and Ca(Fe$_{1-x}$Co$_x$)$_2$As$_2$\cite{kamihara-2008,mu-2008}, 
a unidirectional spin-density-wave has been found. It has been suggested\cite{fang-2008} 
that the undoped system LaOFeAs and CaFe$_2$As$_2$ may have a high-temperature nematic phase and that quantum 
phase transitions also occur as a function of fluorine doping\cite{xu-2008,chuang-2010}.  
This suggests that many of the ideas and results that we present here may be relevant 
to these still poorly understood materials.

The existence of stripe-ordered phases is well established in other complex oxide materials, 
particularly the manganites and the nickelates. In general, these materials tend to be 
``less quantum  mechanical'' than the cuprates in that they are typically insulating 
(although with interesting magnetic properties) and the observed charge-ordered phases 
are very robust. These materials typically have larger electron-phonon interactions and electronic 
correlation are comparatively less dominant in their physics. For this reason they tend to be 
``more classical'' and less prone to quantum phase transitions. However, at least at the classical level, 
many of the issues we discussed above, such as the role of phase separation and Coulomb interactions, 
also play a key role\cite{dagotto-2001}. The thermal melting of a stripe state to a nematic has been 
seen in the manganite material Bi$_x$Ca$_x$MnO$_3$\cite{rubhausen-2000}.

\subsection{Conventional CDW materials}
\label{sec:cdw}

CDWs have been extensively studied since the mid-seventies and there are extensive reviews on their 
properties\cite{gruner-1988,gruner-1994}. From the symmetry point of view there is no difference 
between a CDW and a stripe (or electron smectic). CDW states are usually observed in systems which 
are not particularly strongly correlated, such as the quasi-one-dimensional and quasi-two-dimensional 
dichalcogenides, and the more recently studied tritellurides. These CDW states are reasonably well described as Fermi
liquids (FL) which undergo a CDW transition, commensurate or incommensurate,  
triggered by a nesting condition of the FS\cite{mcmillan-1975,mcmillan-1976}. 
As a result, a part or all of the FS is gapped in which case the CDW may or may 
not retain metallic properties. Instead, in a strongly correlated stripe state, 
which has the same symmetry breaking pattern, at high energy has Luttinger liquid 
behavior\cite{kivelson-1998,emery-2000,carlson-2004}.

What will interest us here is that conventional quasi-2D dichalcogenides, 
the also quasi-2D tritellurides and other similar CDW systems can quantum melt as a 
function of pressure in {\TiSe2}\cite{snow-2003}, or by chemical intercalation as in 
Cu$_x${\TiSe2}\cite{morosan-2006,barath-2007} and Nb$_x$TaS$_2$\cite{lieber-1991}. 
Thus, CDW phases in chalcogenides can serve as a weak-coupling version of the 
problem of quantum melting of a quantum smectic. 
Interestingly, there is strong experimental evidence that both {\TiSe2}\cite{snow-2003} 
and Nb$_x$TaS$_2$\cite{lieber-1991} do not melt directly to an isotropic Fermi fluid 
but go instead through an intermediate, possibly hexatic, phase.\footnote{Cu$_x${\TiSe2} is known to become 
superconducting\cite{morosan-2006}. The temperature-pressure phase diagram of {\TiSe2} exhibits a superconducting dome enclosing the quantum critical point at which the CDW state melts.\cite{kusmartseva-2009}} 
Whether or not the intermediate phases are anisotropic is not known as no transport data is yet available in the 
relevant regime. 

The case of the CDWs in tritellurides is more directly relevant to the theory we will present here. 
Tritellurides are quasi-2D materials which for a broad range of temperatures exhibit a unidirectional 
CDW ({\it i.e.\/} an electronic smectic phase) and whose anisotropic behavior appears to be primarily 
of electronic origin\cite{brouet-2004,laverock-2005,sacchetti-2007,fang-2007}. 
However, the quantum melting of this phase has not been observed yet. 
Theoretical studies have also suggested that it may be possible to have a 
quantum phase transition to a state with more than one CDW in these materials\cite{yao-2006}.

\section{Theories of Stripe Phases}
\label{sec:theories}

\subsection{Stripe Phases in Microscopic Models}
\label{sec:microscopic}

Of all the electronic liquid crystal phases, stripe states have been studied most. There are in fact a number of excellent reviews on this 
topic\cite{carlson-2004,kivelson-2003,vojta-2009} covering both the phenomenology and microscopic mechanisms.  I will only give a brief summary of 
important results and refer to the literature for details. 

As we noted in section \ref{sec:OP-symmetries}, stripe and CDW (and SDW) phases have the same order parameter as they correspond to the same 
broken symmetry state, and therefore the same order parameter $\rho_{\vec{K}}$ (and ${\vec{S}}_{\vec{Q}}$).
\footnote{In principle the order parameter of the stripe state may not be pure sinusoidal and will have higher harmonics of the fundamental 
order parameter.} There is however a conceptual difference. CDW and SDW are normally regarded as weak coupling instabilities of a Fermi liquid 
(or free fermion state) typically triggered by a nesting condition satisfied by the ordering wave vector.\cite{mcmillan-1975,gruner-1988} In this context, the 
quasiparticle spectrum is modified by the partial opening of gaps and a change in the topology of the original Fermi surface (or, equivalently, by the 
formation of ``pockets''). Because of this inherently weak coupling physics, the ordering wave vector is rigidly tied to the Fermi wave vector $k_F$. 

In one-
dimensional systems, non-linearities lead to a more complex form of density wave order, a lattice of solitons, known in this context as 
discommensurations\cite{mcmillan-1976,brazovskii-1984} whose ordering wavevector is no longer necessarily tied to $k_F$. 
A stripe state is essentially a two-dimensional generally 
incommensurate ordered state which is an analog of this strong coupling one-dimensional lattice of discommensurations.\cite{kivelson-1993} Thus, in this 
picture, the spin stripe seen in neutron scattering\cite{tranquada-1995} is pictured as regions of antiferromagnetic (commensurate) order separated by 
anti-phase domain walls (the discommensurations) where the majority of the doped holes reside. This picture is quite hard to achieve by any weak 
coupling approximation such as Hartree-Fock.

Stripe phases were first found in Hartree-Fock studies of Hubbard and $t-J$ models in two dimensions.\cite{zaanen-1989,schulz-1990,poilblanc-1989,machida-1989,kato-1990} 
In  this picture stripe phases are unidirectional charge density waves (CDW) with or without an associated spin-density-wave (SDW) order. As such they 
are 
characterized by a CDW and/or SDW order parameters, $\rho_{\vec{K}}$ and $\vec{S}_{\vec{Q}}$ respectively.\footnote{I will ignore here physically 
correct but more complex orders such as helical phases.}  A Hartree-Fock theory of stripe phases was also developed in the context of the 2DEG in large 
magnetic fields\cite{moessner-1996,koulakov-1996,macdonald-2000} to describe the observed and very large transport anisotropy we discussed above.

As it is usually the case, a serious limitation of the Hartree-Fock approach is that it is inherently reliable only at weak coupling, and hence away from the 
regime of strong correlation of main interest. In particular, all  Hartree-Fock treatments the stripe ground state typically produce an ``empty stripe state'', an 
insulating crystal and therefore not a metallic phase. Thus, in this approach a conducting (metallic) stripe phase can only arise from some sort of quantum 
melting of the insulting crystal and hence not describable in mean-field theory. The phenomenological significance of stripe phases was emphasized by 
several authors, particularly by Emery and Kivelson\cite{emery-1993,kivelson-1996,emery-1999}. 

Mean field theory predicts that, at a fixed value of a electron (or hole) density (doping), the  generally incommensurate ordering wave vectors satisfy the 
relation $\vec{K}=2\vec{Q}$. That this result should generally hold follows from a simple Landau-Ginzburg (LG) analysis of stripe phases 
(see, e.g. \cite{pryadko-1999,berg-2009b} where it is easy to see that a trilinear term of the form 
$\rho^*_{\vec{K}} \vec{S}_{\vec{Q}}\cdot \vec{S}_{\vec{Q}}$ (and its complex conjugate)  is generally allowed in the LG free energy. In ar ordered 
state of this type the antiferromagnetic spin  order is ``deformed'' by anti-phase domain walls whose periodic pattern coincides with that of the charge 
order, as suggested by the observed magnetic structure factor of the stripe state first discovered in the cuprate {\LNSCO}.\cite{tranquada-1995} 

This pattern of CDW and SDW orders has suggested the popular cartoon of stripe phases as antiferromagnetic regions separated by narrow ``rivers of 
charge'' at antiphase domain walls. 
The picture of the stripe phase as an array of rivers suggests a description of stripe phases as a quasi-one-dimensional system. 
As we will see in the next subsection, this picture turned out to be quite  useful for the construction of a strong coupling theory of the physics of the stripe 
phase. On the other hand,  it should not be taken literally in the sense that these rivers always have a finite width which does not have to be small 
compared with the stripe period and in many cases they may well be of similar magnitude. Thus, one may regard this phase as being described by narrow 
 1D regions with significant transversal quantum fluctuations in shape (as it was presented in Ref.\cite{kivelson-1998}) or, equivalently, as  quasi-1D 
 regions with a significant transversal width.

An alternative picture of the stripe phases can be gleaned from the $t-J$ model, the strong coupling limit of the Hubbard model. Since in the resulting 
effective model there is no small parameter, the only (known) way to treat it is to extend to $SU(2)$ symmetry of the Hubbard (and Heisenberg) model to 
either $SU(N)$ or $Sp(N)$ and to use the large $N$ expansion to study its properties.\cite{read-1989b,vojta-1999,vojta-2000,vojta-2009} 
In this (large $N$) limit the undoped antiferromagnet typically has a dimerized ground state, a periodic (crystalline) pattern of valence bond spin singlets. 
Since all degrees of freedom are bound into essentially local singlets this state is a quantum paramagnet. However, it is also ``striped'' in the sense that 
the valence bond crystalline state breaks at least the point group symmetry $\mathcal{C}_4$ of the square lattice as well as translation invariance: it is 
usually a 
period $2$ columnar state.\footnote{This state is a close relative of the resonating valence bond (RVB) state originally proposed as a model system for a 
high $T_c$ superconducting state\cite{anderson-1987,kivelson-1987}, a (non-resonating) VB state.} In the doped system the valence bond crystal 
typically becomes a non-magnetic incommensurate insulating system. Mean-field analyses of these models\cite{vojta-2000} also suggest the existence of 
superconducting states, some of which are ``striped''. Similar results are suggested by variational wave functions based of the 
RVB state.\cite{capello-2008,himeda-2002,yamase-2007b} We should note, however, that mean-field states are no longer controlled by a small 
parameter, such as $1/N$, and hence it is unclear how reliable they may by at the physically relevant case $N=2$. For a detailed (and up-to-date) review 
of this approach see Ref.\cite{vojta-2009}.

There are also extensive numerical studies of stripe phases in Hubbard type models. The best numerical data to date is the density matrix renormalization 
group (DMRG) work of White and Scalapino (and their collaborators) on Hubbard and $t-J$ ladders of various widths (up to 5) and varying particle 
densities\cite{white-1997,white-1998a,white-1998b,white-2000} and by Jackelmann {\it et al} in fairly wide ladders (up to $7$).\cite{jackelmann-2005} An 
excellent summary and discussion  on the results from various numerical results (as well as other insights) can be found in Ref.\cite{carlson-2004}. The 
upshot of all the DMRG work is that there are strong stripe correlations in Hubbard and $t-J$ models which may well be the ground state.\footnote{A 
difficulty in interpreting the DMRG results lies in the boundary conditions that are used that tend to enhance inhomogeneous, stripe-like, phases.}

Much of the work on microscopic mechanisms of stripe formation has been done in models with short range interactions such as the Hubbard and $t-J$ 
models. As it is known,\cite{kivelson-1990,emery-1990,dagotto-1994} models of this type have a strong tendency to {\em electronic phase separation}. As 
we noted in the introduction, the physics of phase separation is essentially the disruption of the correlations of the the Mott (antiferromagnetic) state by the 
doped holes which leads to an effective {\em attractive} interaction among the charge carriers. When  these effects overwhelm the stabilizing effects of the 
Fermi pressure ({\it i.e.\/} the fermion kinetic energy), phase separation follows. In more realistic models, however, longer range (and even Coulomb) 
interactions must be taken into account which tend to frustrate this tendency to phase separation\cite{emery-1993}, as well as a more complex electronic 
structure.\cite{emery-1987} The structure of actual stripe phases in high $T_c$ materials results from a combination of these effects. One of the (largely) 
unsolved questions is the relation between the stripe period and the filling fraction of each stripe at a given density. Most simple minded calculation yield 
simple commensurate filling fractions for each stripe leading to insulating states. At present there are no controlled calculations of these effects, although 
variational estimates have been published\cite{lorenzana-2002}, except for results from DMRG studies on wide ladders.\cite{jackelmann-2005}

\subsection{Phases of Stripe States}
\label{sec:stripe-phases}

We will now discuss the strong coupling picture of the stripe 
phases.\cite{emery-2000,vojta-2000,granath-2001,arrigoni-2004,carlson-2004,kivelson-2007}. We will assume that a stripe phase exists with a 
fixed (generally incommensurate) 
wave vector $\vec{K}$ and a fixed filling fraction (or density) on each stripe. In this picture a stripe phase is equivalent to an array of ladders of certain 
width. In what follows we will assume that each stripe has a finite {\em spin gap}: a {\em Luther-Emery liquid}.\cite{emery-1979,luther-1974} 

\subsubsection{Physics of the 2-leg ladder}
\label{sec:2leg}

The assumption of the existence of a finite spin gap in ladders can be justified in several ways. In DMRG studies of Hubbard 
and $t-J$ ladders in a rather broad density range, $0<x<0.3$, it is found that the ground state has a finite spin gap.\cite{noack-1997} Similar results were 
found analytically in the weak coupling regime.\cite{balents-1996,lin-1997,lin-1998,zachar-1999}

Why there is a spin gap? There is actually a very simple argument for it.\cite{emery-1997}
In the non-interacting limit, $U=V=0$, the two-leg ladder has two bands with two different Fermi wave vectors, $p_{F1}\neq p_{F2}$.
 Let us consider the effects of interactions in this weak coupling regime.
  The only allowed processes involve an {\em even} number of electrons.
  In this limit is is easy to see that the coupling of CDW fluctuations with $Q_1=2p_{F1} \neq Q_2=2p_{F2}$ is suppressed due to the mismatch of their 
  ordering wave vectors. In this case, scattering of electron pairs with zero center of mass momentum from one system to the other is a peturbatively 
  (marginally) relevant interaction. The spin gap arises since the electrons can gain zero-point energy by delocalizing between the two bands.  
  To do that, the electrons need to pair, which may cost some energy.  
When the energy gained by delocalizing between 
the two bands exceeds the energy cost of pairing, the system is driven to 
a spin-gap phase. 

This physics is borne out by detailed numerical (DMRG) calculations, even in systems with only repulsive interactions. Indeed, 
at $x=0$ (the undoped ladder) the system is in a Mott insulating state, with a unique fully gapped ground state (``$C0S0$'' in the language of 
Ref.\cite{balents-1996}). In the strong coupling limit (in which the rungs of the ladder are spin singlet valence bonds), $U\gg t$, the spin gap is large: 
$\Delta_s \sim J/2$.\cite{troyer-1995} 

At low doping,  $0<x<x_c \sim 0.3$, the doped ladder is in a Luther-Emery liquid: there is no charge gap and large spin gap (``$C1S0$''). In fact, in this 
regime the spin gap is found to decrease monotonically as doping increases,
$\Delta_s \downarrow$ as $x \uparrow$, and vanishes at a critical value $x_c$: $\Delta_s \to 0$ as $x \to x_c$.

The most straightforward way to describe this system is to use bosonization. Although in the ladder system has several bands of  electrons that have 
charge and spin degrees of freedom, in the low energy regime the effective description is considerably simplified. Indeed, in this regime it is sufficient to 
consider only one effective bosonized charge field and one bosonized spin field. Since there is a spin gap, $\Delta_s \neq 0$, the spin sector is massive. 
In contrast, the charge sector is only massive at $x=0$, where there is a finite Mott gap $\Delta_M$. 

The effective Hamiltonian for the charge degrees of freedom in this (Luther-Emery) phase is
  \beq
  H=\int dy \frac{v_c}{2} \left[ \frac{1}{K} \left(\partial_y \theta\right)^2+K \left(\partial_x \phi\right)^2\right]+\ldots
  \eeq
  where $\phi$ is the CDW phase field, and $\theta$ is the SC phase field. They satisfy canonical commutation relations 
  \beq
  [\phi(y^\prime),\partial_y \theta(y)]=i\delta(y-y^\prime)
  \eeq
The parameters of this effective theory, the spin gap $\Delta_s$, the charge Luttinger parameter $K$, the charge velocity $v_c$, and the chemical 
potential $\mu$, have non-universal but smooth dependences on the doping $x$ and on the parameters of the microscopic Hamiltonian, the hopping 
matrix elements $t^\prime/t$ and the Hubbard interaction $U/t$. The ellipsis 
$\ldots$ in the effective Hamiltonian represent cosine potentials responsible for the Mott gap $\Delta_M$ in the undoped system ($x=0$).
It can be shown that the spectrum in the low doping regime, $x\to 0$, consists of gapless and spinless charge $2e$ fermionic solitons.
  
The charge Luttinger parameter is found to approach $K \to 1/2$ as $x\to 0$. As $x$ increases, so does $K$ reaching the value 
$K \sim 1$ for $x\sim 0.1$. On the other hand $K\sim2$ for $x\sim x_c$ where the pin gap vanishes. The temperature dependence of the superconducting 
and CDW susceptibilities have the scaling behavior
 \bea
 \chi_{\rm SC}&\sim& \frac{\Delta_s}{T^{2-K}} \\
 \chi_{\rm CDW}&\sim& \frac{\Delta_s}{T^{2-K^{-1}}}
 \eea
Thus, both susceptibilities diverge $\chi_{\rm CDW}(T)\to \infty$ and $\chi_{\rm SC}(T)\to \infty$ for $0<x<x_c$. However,  for $x\lesssim 0.1$, the SC 
susceptibility is more divergent: $\chi_{\rm SC}\gg \chi_{\rm CDW}$. Hence, the doped ladder in the Luther-Emery regime is effectively a 1D 
superconductor even for a system with nominally repulsive interactions ({\it i.e.\/} without ``pairing'').

\subsubsection{The spin-gap stripe state}
\label{sec:spin-gap}

Now consider a system with $N$ stripes, each labeled by an integer
$a=1,\ldots,N$.  We will consider first the phase in which there is a spin
gap. Here, the spin fluctuations are effectively frozen out at low energies.
Nevertheless each stripe $a$ has two degrees of freedom\cite{kivelson-1998}:
a transverse displacement field
which describes the local dynamics
of  the configuration of each stripe, and the phase field $\phi_a$ for the
charge
fluctuations on each stripe. The action of the generalized Luttinger liquid
which describes
the smectic charged fluid of the stripe state is obtained by integrating out
the local
shape fluctuations associated with the displacement fields. These fluctuations
give rise
to a finite renormalization of the Luttinger parameter and velocity of each
stripe. More
importantly, the shape fluctuations, combined with the long-wavelength
inter-stripe
Coulomb interactions, induce inter-stripe density-density and current-current
interactions, leading to an imaginary time Lagrangian density of the form
\begin{equation}
{\cal L}_{\rm smectic}=\frac{1}{2}\; \sum_{a, a',\mu}  \;
j_{\mu}^a(x) \; \tilde W_{\mu}(a - a') \; j_{\mu}^{a'}(x).
\label{eq:smectic}
\end{equation}
These operators
are {\sl marginal}, {\it i.e.} have scaling dimension $2$, and preserve
the {\sl smectic symmetry}  $\phi_a \to \phi_a+ \alpha_a$ (where $\alpha_a$
is constant on each stripe) of the decoupled Luttinger fluids. Whenever this
symmetry is
exact, the charge-density-wave order parameters of the individual stripes do
not lock with
each other, and the charge density profiles on each stripe can slide relative
to each other
without an energy cost. In other words, there is no rigidity to shear
deformations of the
charge configuration on nearby stripes.
This is the {\sl smectic metal} phase\cite{kivelson-1998}, a {\em sliding Luttinger liquid}.\cite{vishwanath-2001}

The fixed point action for a generic smectic metal phase
thus has the form (in Fourier space)
\begin{eqnarray}
S&&=\sum_Q {\frac{K(Q)}{2}}  \left\{ \frac{\omega^2 }{v(Q)}+ v(Q)
k^2 \right\} |\phi(Q)|^2
\nonumber \\
&&=\sum_Q {\frac{1}{2K(Q)}} \left\{
{\frac{\omega^2}{v(Q)}}+ v (Q) k^2
\right\}
|\theta(Q)|^2
\label{eq:action-general}
\end{eqnarray}
where $Q=(\omega,k,k_\perp)$, and $\theta$ is the field {\sl dual} to
$\phi$. Here $k$ is the momentum {\sl along} the stripe and $k_\perp$
perpendicular to the stripes.
The kernels $K(Q)$ and $v(Q)$ are analytic
functions of $Q$ whose form depends on microscopic details, {\it e.\ g.\/}
at weak coupling they are functions of the inter-stripe Fourier transforms of
the
forward and backward scattering amplitudes
$g_2(k_\perp)$ and $g_4(k_\perp)$,
respectively. In practice, up to irrelevant operators, it is sufficient to keep the dependence of the kernels only on the transverse momentum $k_\perp$.
Thus, the smectic fixed point is characterized by the effective Luttinger parameter and velocity (functions),
$K(k_\perp)$ and
$v(k_\perp)$. Much like the ordinary 1D Luttinger
liquid, this ``fixed point'' is characterized by power-law decay of correlations functions. This effective field theory also yields the correct low energy description of the quantum Hall stripe phase of the 2DEG in large magnetic fields.\cite{fradkin-1999,fertig-1999,macdonald-2000,barci-2002a,lawler-2004}

In the presence of a spin gap, single electron tunneling is
irrelevant\cite{emery-1997},
and
the only potentially relevant interactions involving pairs of stripes $a, a'$
are singlet pair (Josephson) tunneling, and the coupling between the CDW order
parameters.
These interactions have the form
${\cal H}_{\rm int}=\sum_n \left({\cal H}_{\rm SC}^n+{\cal H}_{\rm CDW}^n
\right)$
for $a'-a=n$, where
\begin{eqnarray}
{\cal H}_{\rm SC}^n=&&
\left({\frac{\Lambda}{2\pi}}\right)^2\sum_{a} {\cal J}_n
\cos [\sqrt{2 \pi}  (\theta_a -
\theta_{a+n})]
\nonumber \\
{\cal H}_{\rm CDW}^n=&&\left({\frac{\Lambda}{2\pi}}\right)^2\sum_{a}
 {\cal V}_n \cos [\sqrt{2 \pi} (\phi_a - \phi_{a+n})].
\label{eq:Hint}
\end{eqnarray}
Here ${\cal J}_n$ are the inter-stripe Josephson couplings (SC), $ {\cal V}_n$
are the $2k_F$ component of the inter-stripe density-density (CDW)
interactions,
and $\Lambda$ is an ultra-violet cutoff, $\Lambda\sim 1/a$
where $a$ is a lattice constant.
A straightforward calculation, yields the scaling dimensions
$\Delta_{1,n}\equiv\Delta_{{\rm SC},n}$ and
$\Delta_{-1,n}\equiv\Delta_{{\rm CDW},n}$
of ${\cal H}_{\rm SC}^n$ and ${\cal H}_{\rm CDW}^n$:
\bea
\Delta_{\pm 1,n}=
\int_{-\pi}^\pi {\frac{d k_\perp}{2\pi}} \;
\left[\kappa(k_\perp)\right]^{\pm 1} \left(1-\cos nk_\perp \right) ,
\label{eq:dimensions1}
\eea
where $\kappa(k_\perp)\equiv K(0,0,k_\perp)$.
Since $\kappa(k_\perp)$ is a periodic
function of $k_\perp$ with period $2\pi$, $\kappa(k_\perp)$ has a convergent
Fourier
expansion of the form $\kappa(k_\perp)=\sum_n \kappa_n \cos n k_\perp$. We will
parametrize the fixed point theory by the coefficients $\kappa_n$, which are
smooth
non-universal functions. In what follows we shall discuss
the behavior of the simplified model  with $\kappa(k_\perp)= \kappa_0+
 \kappa_1 {\cos
k_\perp}$. Here, $ \kappa_0$  can be thought of as the intra-stripe inverse
Luttinger
parameter, and $ \kappa_1$ is a measure of the nearest neighbor inter-stripe
coupling.
For stability we require $ \kappa_0 > \kappa_1$.

Since it is unphysical to consider longer range interactions in $H_{int}$ than
are present in the fixed point Hamiltonian, we treat only
perturbations with
$n=1$, whose dimensions are
 \beq
 \Delta_{{\rm SC},1}\equiv \Delta_{\rm SC}=\kappa_0-{\frac{\kappa_1}{2}}
 \eeq
 and
\beq
\Delta_{{\rm CDW},1}\equiv \Delta_{\rm CDW} =\frac{2}{\left(\kappa_0-\kappa_1+
\sqrt{\kappa_0^2-
\kappa_1^2}\right)}
\eeq
For a more general function $\kappa(k_\perp)$,
operators with larger $n$ must also be considered, but the results are
qualitatively unchanged \cite{ohern-1999,vishwanath-2001}.\footnote{$\Delta_{{\rm SC},2}$ is the most relevant
operator. For a model with
$\kappa(k_{\perp})=[\kappa_0+\kappa_1\cos(k_{\perp})]^2$, 
all perturbations are
irrelevant
for large $\kappa_0$ and small $|\kappa_0-\kappa_1|$.}
\begin{figure}[t!]
\begin{center}
\includegraphics[width=0.6\textwidth]{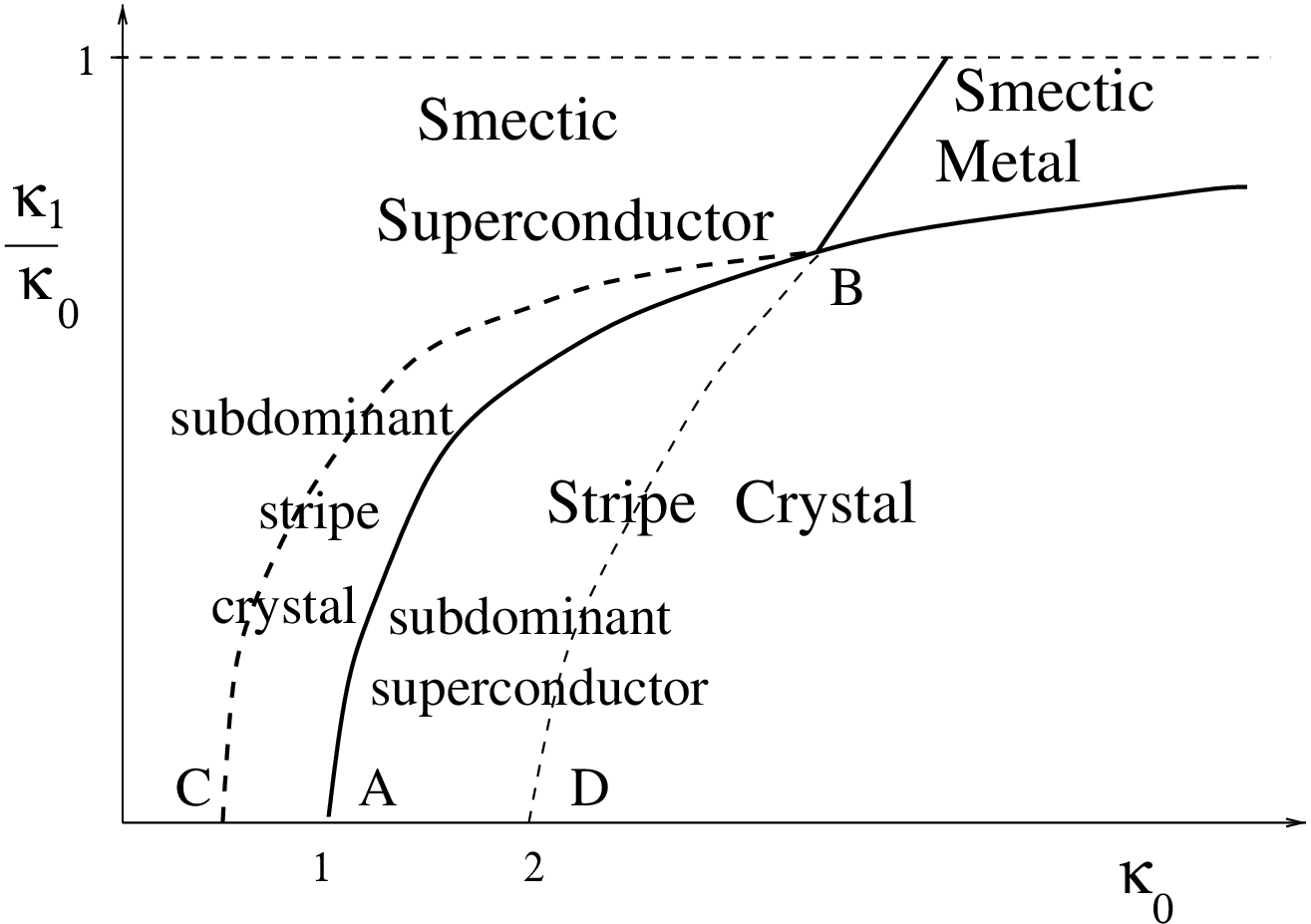}
\end{center}
\caption
{Phase diagram for a stripe state with a spin gap.}
\label{fig:spin-gap}
\end{figure}

In Fig. \ref{fig:spin-gap} we present the phase diagram of this model.
The dark $AB$ curve is the set of points where
$\Delta_{{\rm CDW}}=\Delta_{{\rm SC}}$, and it is a line of first order
transitions. To the right of this line the inter-stripe CDW coupling is
the most relevant perturbation, indicating an instability of
the system to the formation of a 2D stripe crystal\cite{kivelson-1998}.
To the left, Josephson tunneling (which still preserves the smectic symmetry)
is the most relevant, so this phase is a 2D smectic superconductor. (Here we
have neglected the possibility of coexistence since a first order transition
seems more likely). Note that there is a region of $ \kappa_{0} \geq 1$, and
large enough $ \kappa_{1}$, where the global order is superconducting
although, in the absence of inter-stripe interactions (which roughly corresponds
to
$\kappa_{1}=0$), the superconducting 
 fluctuations are subdominant.  There is also a (strong
coupling) regime above the curve $CB$ where {\sl both} Josephson tunneling
{\sl and} the CDW
coupling {\sl are irrelevant} at low energies.
Thus, in this regime {\sl the smectic metal
state is stable}. This phase is a 2D smectic non-Fermi liquid in which there
is coherent transport {\sl only} along the stripes.

To go beyond this description we need to construct an effective theory of the {\em two-dimensional} ordered phase. For instance, the superconducting 
state is a 2D striped superconductor, whereas the crystal is a bidirectional charge density wave. A theory of these 2D ordered phases can be developed 
by combining the quasi-one-dimensional renormalization group with an effective inter-stripe mean field theory, as in Ref.\cite{carlson-2000}, which in turn 
can be fed into a 2D renormalization group theory.\cite{affleck-1996} One advantage of this approach is that the inter-stripe mean field theory has the 
same analytic structure as the dimensional crossover RG (see Ref.\cite{arrigoni-2004}).

Let us consider the superconducting state, a {\em striped} superconductor. In the way we constructed this state all ladders are {\em equivalent}. Hence this 
is a period $2$ stripe (columnar) SC phase, similar to the one discussed by Vojta.\cite{vojta-2009} Let us use inter-stripe mean field theory to estimate the 
critical temperature of the 2D state. For the isolated ladder, $T_c=0$ as required by the Mermin-Wagner theorem.
If the inter-stripe Josephson and CDW couplings are non-zero,
  ${\mathcal J}\neq 0$ and ${\mathcal V}\neq 0$, the system will now have a finite SC critical temperature, $T_c >0$.
Now, for
$x \lesssim0.1$, CDW couplings are irrelevant as in this range $1/2<K<1$. Hence, in the same range, the 
inter-ladder Josephson coupling are relevant and 
lead to a SC state in a small $x$ with a somewhat low $T_c$ which, in inter-stripe (or `chain')  mean field theory can be estimated by
 \beq
 2 {\mathcal J} \chi_{\rm SC}(T_c)=1
 \eeq
In this regime, however, $T_c \propto \delta t \; x$ and it is low due to the low carrier density.
Conversely, for larger $x$, $K>1$ and $\chi_{CDW}$  is more strongly divergent than $\chi_{SC}$. Thus, for $x\gtrsim 0.1$ the CDW couplings become 
more relevant. This leads to an insulating incommensurate CDW state with ordering wave number  $P=2\pi  x$.  

In the scenario we just outlined\cite{arrigoni-2004,kivelson-2007} in the 2D regime the system has a {\em first order} transition from a superconducting 
state to a non-superconducting  phase with charge order. However at large enough inter-stripe forward scattering interactions both couplings become 
irrelevant and there is a quantum {\em bicritical} point separating both phases from a smectic metal (as depicted in Fig.\ref{fig:spin-gap}). However, an 
alternative possibility is that instead of a bicritical point, we may have a quantum {\em tetracritical} point and a phase in which SC and CDW orders 
coexist. 

\section[Optimal Inhomogeneity]{Is Inhomogeneity Good or Bad for Superconductivity?}
\label{sec:optimal}

The analysis we just did raises the question of whether stripe order (that is, some form of spatial charge inhomogeneity) is good or bad for superconductivity. This question was addressed in some detail in Refs.\cite{arrigoni-2004,kivelson-2007} where it was concluded that a) there is an {\em optimal degree of inhomogeneity at which $T_c$ reaches a maximum} and b) that charge order in a system with a spin gap can provide a mechanism of ``high temperature superconductivity'' (the meaning of which we will specify below).

\begin{figure}[t!]
\psfrag{t}{$t$}
\psfrag{dt}{$\delta t$}
\psfrag{e}{$\epsilon$}
\psfrag{-e}{$-\epsilon$}
\psfrag{tp}{$t^\prime$}
\psfrag{a}{$A$}
\psfrag{b}{$B$}
\begin{center}
\includegraphics[width=0.5\textwidth]{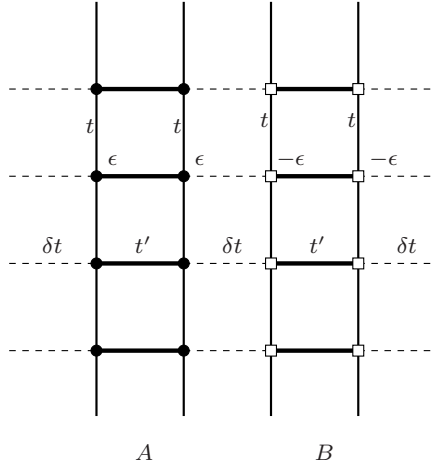}
\end{center}
\caption{Model of a period 4 stripe phase}
\label{fig:period4}
\end{figure}

The argument goes as follows. Consider a system with a {\em period $4$} stripe phase, consisting of an
alternating array of inequivalent  $A$ and $B$ type ladders in the Luther-Emery regime.\footnote{In Ref.\cite{granath-2001} a similar pattern was also considered except that the (say) `$B$' stripes do no have a spin gap. This patterns was used to show how a crude model with nodal quasiparticles can arise in an inhomogeneous state.}  The inter-stripe mean field theory estimate for the superconducting and CDW critical temperatures now takes the somewhat more complex form:
\beq
(2{\mathcal J})^2 \chi_{\rm SC}^A(T_c) \chi_{\rm SC}^B(T_c)=1
\eeq
for the superconducting $T_c$, and 
 \beq
 (2{\mathcal V})^2 \chi_{\rm CDW}^A(P,T_c) \chi_{\rm CDW}^B(P,T_c)=1
 \eeq
 for the CDW $T_c$. In particular, the
2D CDW order is greatly suppressed due to the 
mismatch  between ordering vectors, $P_A$ and $P_B$, on neighboring ladders

For inequivalent $A$ and $B$ ladders SC beats CDW if the corresponding Luttinger parameters satisfy the inequalities
\beq
2 > K_A^{-1}+K_B^{-1} - K_A; \ \ \ 2 > K_A^{-1}+K_B^{-1} - K_B
\eeq
The SC critical temperature is then found to obey a {\em power law} scaling form (instead of the essential singularity of the BCS theory of superconductivity):
\beq
T_c \sim \Delta_s \left(\frac {\cal J} {\widetilde W} \right)^{\alpha}; \   
\alpha=\frac {2K_AK_B}{[4K_AK_B-
K_A-K_B]}
\eeq
A simple estimate of the effective inter-stripe Josephson coupling, ${\cal J}\sim \delta t^2/J$ and of the high energy scale $\widetilde W\sim J$, implies that the superconducting critical temperature $T_c$ is (power law) small  
for small ${\cal J}$!, with an exponent that typically is $\alpha  \sim 1$.

\begin{figure}[h!]
\psfrag{Tc}{ $T_c$}
\psfrag{Ds}{$\Delta_s(x)$}
\psfrag{Kc}{$K_c$}
\psfrag{2}{$2$}
\psfrag{1}{$1$}
\psfrag{0}{ $0$}
\psfrag{1/2}{$\frac{1}{2}$}
\psfrag{a}{$\Delta_s(2) \left(\displaystyle{\frac{\delta t}{J}}\right)^{\frac{4}{3}}$}
\psfrag{b}{$\Delta_s(1) \left(\displaystyle{\frac{\delta t}{J}}\right)$}
\psfrag{SC}{SC}
\psfrag{CDW}{CDW}
\psfrag{x}{$x$}
\psfrag{xc}{$x_c$}
\psfrag{xc(2)}{$x_c(2)$}
\psfrag{xc(4)}{$x_c(4)$}
\psfrag{x1}{$x_1$}
\psfrag{J}{$\displaystyle{\frac{J}{2}}$}
\begin{center}
\includegraphics[width=0.7\textwidth]{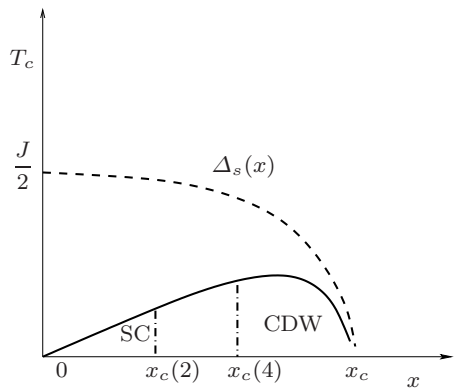} 
\end{center}
\label{fig:optimal}
\caption{Evolution of the superconducting critical temperature with doping.}
\end{figure}

These arguments can be used to sketch a phase diagram of the type presented in Fig.\ref{fig:optimal} which shows the qualitative dependence of the SC 
$T_c$ with doping $x$. The broken line shown is the spin gap $\Delta_s(x)$ as a function of doping $x$ and, within this analysis, it must be an upper 
bound on $T_c$. Our arguments then showed that a period 4 structure can have a substantially larger $T_c$ than a period 2 stripe. Consequently, the 
critical dopings, $x_c(2)$ and $x_c(4)$,  for the SC-CDW quantum phase transition must move to higher values of $x$ for period 4 compared with period 
2. On the other hand, for $x \gtrsim x_c$ the isolated ladders
do not have a spin gap, and this strong coupling mechanism is no longer operative.

How reliable are these estimates? What we have are mean-field estimates for $T_c$ and it is an upper bound to the actual $T_c$.
As it is usually the case, $T_c$ should be suppressed by phase fluctuations by up to a factor of 2. 
On the other hand, perturbative RG studies for small ${\cal J}$
yield the {\em same power law dependence}.
This result is asymptotically exact for ${\cal J}<<\widetilde W$.
Since $T_c$ is a smooth function of $\delta
t/{\cal J}$, it is reasonable to extrapolate for $\delta t \sim{\cal J}$. 
Hence, $T_c^{\rm max} \propto \Delta_s$ and we have a ``high $T_c$''.
This is 
in contrast to the exponentially small $T_c$ obtained in a BCS-like mechanism.

Now, having convinced ourselves that a period 4 stripe will have a larger SC $T_c$ than a period 2 stripe one may wonder if an even longer period stripe 
state would do better. It is easy to see that there will be a problem with this proposal. Clearly, although the argument we just presented would suggest that 
the exponents will also be of order $1$ for longer periods, the problem now is that the effective couplings become very small very quickly as the 
Josephson coupling has an {\em exponential} dependence on distance (tunneling!) . Thus, there must be an optimal period for this mechanism and it is 
likely to be a number larger than 2 but smaller than (say) 6.

In summary, we have shown that in systems with strong repulsive interactions (and without attractive interactions), an (inhomogeneous) stripe-ordered 
state can support a 2D superconducting state with a high critical temperature, in the sense that it is not exponentially suppressed, with a high paring scale 
(the spin gap). This state is an inhomogeneous version of the RVB mechanism.\cite{anderson-1987,kivelson-1987,lee-2006} The arguments suggest that 
there is an optimal degree of inhomogeneity. There is suggestive evidence in ARPES data in {\LBCO} that show a large pairing scale in the stripe-
ordered state which support this picture.\cite{valla-2006,he-2008}

\section{The Striped Superconductor: a Pair Density Wave state}
\label{sec:pdw}

We now turn to a novel type of striped superconductor, the pair density wave state. Berg {\it et al}\cite{berg-2007,berg-2008a,berg-2009b} have 
recently proposed this state as a symmetry-based explanation of the spectacular dynamical layer decoupling seen in stripe-ordered {\LBCO} (and 
{\LNSCO})\cite{li-2007,tranquada-2008,hucker-2009}, and in {\LSCO} in magnetic fields.\cite{schafgans-2008}

Summary of experimental facts for {\LBCO} near $x=1/8$:
\begin{itemize}
\item
ARPES finds an anti-nodal d-wave SC gap that is large and unsuppressed at $x=1/8$. Hence, there is a large pairing scale in the stripe-ordered state.
\item
Resonant X-Ray scattering finds static charge stripe order for $T< T_{charge}=54 K$.
\item
Neutron Scattering finds static Stripe Spin order $T < T_{spin}= 42K$.
\item
The in-plane resistivity $\rho_{ab}$ drops rapidly to zero from $T_{spin}$ to $T_{KT}$ (the Kosterlitz-Thouless (KT) transition).
\item
$\rho_{ab}$ shows KT behavior for $T_{spin} > T > T_{KT}$.
\item
$\rho_{c}$ increases as $T$ decreases for $T>T^{**}\approx  35K$.
\item
$\rho_c \to 0$ as $T \to T_{3D}= 10K$ (the bulk 3D resistive transition).
\item
$\rho_c / \rho_{ab} \to \infty$ for $T_{KT} > T > T_{3D}$.
\item
Theres is a Meissner state only below $T_c= 4K$.
\end{itemize}

How do we understand these remarkable effects that can be summarized as follows: There is a broad temperature range, $T_{3D} < T < T_{2D}$ with 2D superconductivity but not in 3D, as if there is not interlayer Josephson coupling.
In this regime there is both striped charge and spin order.
This can only happen if there is a special symmetry of the superconductor in the striped state that leads to an almost complete cancellation of the c-axis Josephson coupling.

What else do we know? The stripe state in the LTT (``low temperature tetragonal'') crystal structure of {\LBCO} has two planes in the unit cell.
Stripes in the 2nd neighbor planes are shifted by half a period to minimize the Coulomb interaction: 4 planes per unit cell.
The anti-ferromagnetic spin order suffers a $\pi$ phase shift accross the charge stripe which has period 4.
Berg {\it et al}\cite{berg-2007} proposed that the superconducting order is also striped and also suffers a $\pi$ phase shift.
The superconductivity resides in the spin gap regions and there is a $\pi$ phase shift in the SC order across the anti-ferromagnetic regions.

The PDW SC state has {\em intertwined} striped charge, spin and superconducting orders. \footnote{While there is some numerical evidence for a state of this type  in variational Monte Carlo calculations \cite{himeda-2002} and in slave particle mean field theory\cite{raczkowski-2007,capello-2008} (see, however, Ref.\cite{yang-2008b} and \cite{loder-2009}), a consistent and controlled microscopic theory is yet to be developed. Since the difference between the energies of the competing states seen numerically is quite small one must conclude that they are all reasonably likely.}

\begin{figure}[hbt]
\begin{center}
\subfigure{\includegraphics[width=0.54\textwidth]{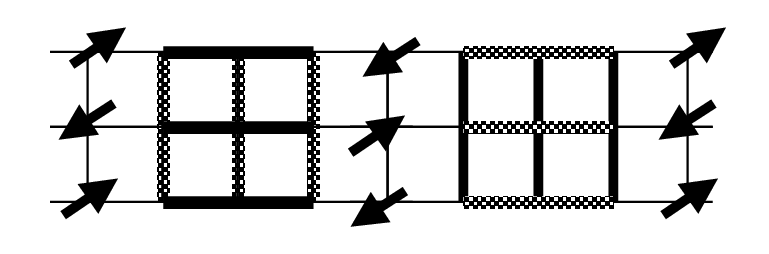}}
\subfigure{\includegraphics[width=0.44\textwidth]{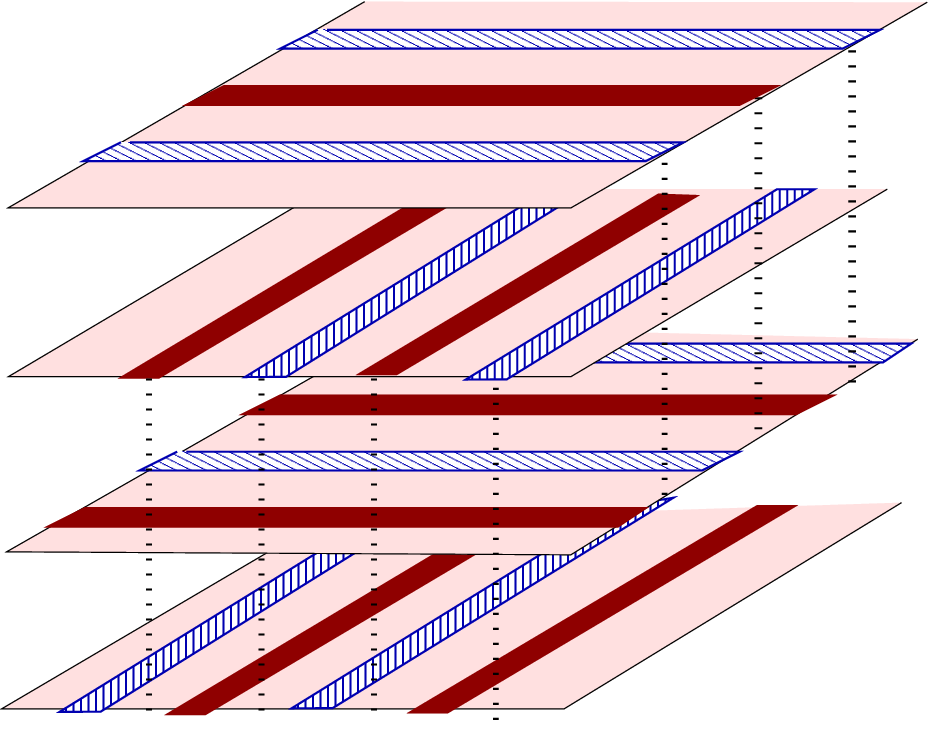}}
\caption{Period 4 Striped Superconducting State}
\end{center}
\end{figure}

How does this state solve the puzzle? If this order is perfect,  the Josephson coupling between neighboring planes cancels exactly due to the symmetry of the periodic array of $\pi$  textures, {\it i.e.\/} the spatial average of the SC order parameter is exactly zero.
The Josephson couplings $J_1$ and  $J_2$ between planes two and three layers apart also cancel by symmetry. 
The first non-vanishing coupling $J_3$ occurs at four spacings. It is quite small and it is responsible for the non-zero but very low $T_c$
Defects and/or discommensurations gives rise to  small Josephson coupling  $J_0$  neighboring planes.

Are there other interactions? It is possible to have an inter-plane biquadratic coupling involving the product SC of the order parameters between neighboring planes  $\Delta_1 \Delta_2$ and the product of spin stripe order parameters also on neighboring planes $\vec{M}_1 \cdot \vec{M}_2$.
However in the LTT structure $\vec{M}_1 \cdot \vec{M}_2=0$ and there is no such coupling.
In a large enough perpendicular magnetic field it is possible (spin flop transition) to induce such a term and hence an effective Josephson coupling.
Thus in this state there should be a strong suppression of the 3D SC  Tc but not of the 2D SC $T_c$. 

On the other hand, away from $x=1/8$ there is no perfect commensuration.
Discommensurations are defects that induce a finite Josephson coupling between neighboring planes $J_1 ~ |x-1/8|^2$, leading to an increase of  the 3D SC Tc away from $x=1/8$.
Similar effects arise from disorder which also lead to a rise in the 3D SC $T_c$. 

\subsection{Landau-Ginzburg Theory of the Pair Density Wave}
\label{sec:GL-PDW}

In what follows we will rely heavily on the results of Refs.\cite{berg-2008a,berg-2009,berg-2009b}. We begin with a description of the 
order parameters:
\begin{enumerate}
\item
PDW (Striped) SC: 
\beq
\Delta(\vec{r})=\Delta_{\vec{Q}}(\vec{r}) e^{i \vec{Q}\cdot \vec{r}} +  \Delta_{-\vec{Q}}(\vec{r}) e^{-i \vec{Q}\cdot \vec{r} } 
\eeq
 complex charge 2e singlet pair condensate with wave vector $\vec{Q}$, (i.e. an FFLO type state at zero magnetic field)\footnote{A state that is usually 
 described as a pair crystal is commonly known as a pair density wave.\cite{chen-2004,tesanovic-2005} However that state cannot be distinguished by 
 symmetry from a (two) CDWs coexisting with a uniform SC.} 
 \item
Nematic: detects breaking of rotational symmetry:  $N$, a real neutral pseudo-scalar order parameter
 \item
Charge stripe:  $\rho_{\vec{K}}$, unidirectional charge stripe with wave vector $\vec{K}$
 \item
Spin stripe order parameter:  $\vec{S}_{\vec{Q}}$, a neutral complex spin vector order parameter.
\end{enumerate}
These order parameters have the following transformation properties under rotations by $\pi/2$, $ \mathcal{R}_{\pi/2}$:
\begin{enumerate}
\item
The nematic order parameter changes sign:  $N \to -N$
\item
The CDW ordering wave vector rotates:  $\rho_{\vec{K}}  \to  \rho_{\mathcal{R}_{\pi/2}\vec{K}}$
\item
The SDW ordering wave vector also rotates: $\vec{S}_{\vec{Q}}  \to  \vec{S}_{\mathcal{R}_{\pi/2}\vec{Q}}$
\item
The striped SC (s or d wave) order parameter: $\Delta_{\pm \vec{Q}} \to \pm \Delta_{\pm \mathcal{R}_{\pi/2}\vec{Q}}$ ($+$ for $s$-wave, $-$ for $d$-
wave)
\end{enumerate}
and by translations by $\vec{R}$
\beq
N \to N ,\quad  \rho_{\vec{K}}  \to e^{i\vec{K}\cdot \vec{R}} \rho_{\vec{K}},  \quad \vec{S}_{\vec{Q}} \to e^{i\vec{Q} \cdot \vec{R}} \vec{S}_
{\vec{Q}}
\eeq

The Landau-Ginzburg free energy functional is, as usual, a sum of terms of the form
\begin{equation}
\mathcal{F}=\mathcal{F}_{2}+\mathcal{F}_{3}+\mathcal{F}_{4}+\ldots
\label{F}
\end{equation}
where $\mathcal{F}_{2}$, the quadratic term, is simply a sum of decoupled
terms for each order parameter. There exist a number of trilinear terms mixing several of the order parameters described above. They are
\begin{eqnarray}
\mathcal{F}_{3} &=&\gamma _{s}[\rho _{-\vec{K}}\vec{S}_{\vec{Q}}\cdot
\vec{S}_{\vec{Q}} + \rho_{-{\bar{\vec{K}}}}\vec{S}_{\bar{\vec{Q}}}\cdot \vec{S}_{\bar{\vec{Q}}}+\mathrm{c.c.}] \\
&&+\gamma _{\Delta }[\rho _{-\vec{K}}\Delta _{-\vec{Q}}^{\star
}\Delta_{\vec{Q}} +\rho_{-\bar{\vec{K}}}\Delta _{-\bar{\vec{Q}}}^{\star }\Delta_{\bar{\vec{Q}}}+\mathrm{c.c.}]  \nonumber \\
&&+g_{\Delta }N[\Delta _{\vec{Q}}^{\star }\Delta _{\vec{Q}}+ \Delta_{-\vec{Q}}^{\star }\Delta _{-\vec{Q}}-\Delta _{\bar{\vec{Q}}}^{\star
} \Delta_{\bar{\vec{Q}}}-\Delta _{-\bar{\vec{Q}}}^{\star}\Delta _{-\bar{\vec{Q}}}]  \nonumber \\
&&+g_{s}N[\vec{S}_{-\vec{Q}}\cdot \vec{S}_{\vec{Q}}-\vec{S}_{-\bar{\vec{Q}}}\cdot \vec{S}_{\bar{\vec{Q}}}]  \nonumber \\
&&+g_{c}N[\rho _{-\vec{K}}\rho _{\vec{K}}-\rho _{-\bar{\vec{K}}}\rho _{\bar{\vec{K}}}],  \nonumber
\label{eq:F3}
\end{eqnarray}
where $\bar{\vec{Q}}=\mathcal{R}_{\pi/2}\vec{Q}$, and $\bar{\vec{K}}=\mathcal{R}_{\pi/2}\vec{K}$.
The fourth order term, which is more or less standard, is shown
explicitly below. 

Several consequences follow directly from the form of the trilinear terms, Eq.\eqref{eq:F3}. One is that, at least in a fully translationally invariant system, 
the first two terms of Eq.\eqref{eq:F3} implies a relation between the ordering wave vectors: $\vec{K}=2\vec{Q}$. Also, as we will see below, these 
terms implies the existence of vortices of the SC order with {\em half} the flux quantum.

Another important feature of the PDW SC is that it implies the existence of a non-zero charge $4e$ {\em uniform} SC state. Indeed, if we denote by $
\Delta_4$ the (uniform) charge $4e$ SC order parameter, then the following term in the LG expansion is allowed
\beq
\mathcal{F}^\prime_3=g_4 \left[\Delta_4^*\left(\Delta_{\vec{Q}}\Delta_{-\vec{Q}}+\textrm{rotation\; by} \frac{\pi}{2}\right)+\textrm{c.c.}\right]
\eeq
Hence, the existence of striped SC order (PDW) implies the existence uniform charge $4e$ SC order!\footnote{A charge $4e$ SC order parameter is an 
expectation value of a four fermion operator.} 

We should also consider a different phase in which there are coexisting uniform and striped SC orders, as it presumably happens at low temperatures in 
{\LBCO}. If this is so, there is a non-zero PDW SC order parameter $\Delta_{\vec{Q}}$ as well as an uniform (d-wave)  SC order parameter $\Delta_0$ 
which are coupled by new (also trilinear) terms in the LG free energy of the form
\beq
\mathcal{F}_{3,u}=\gamma_\Delta \Delta_0^* (\rho_{\vec{Q}} \Delta_{-\vec{Q}}+\rho_{-\vec{Q}} \Delta_{\vec{Q}})+g_\rho \rho_{-2\vec{Q}} 
\rho_{\vec{Q}}^2 +\frac{\pi}{2}\; \textrm{rotation}+ \textrm{c.c.}
\eeq
If $\Delta_0\neq 0$ and $\Delta_{\vec{Q}}\neq 0$, there is a new $\rho_{\vec{Q}}$ component of the charge order!. Also,
the small uniform component $\Delta_0$ removes the sensitivity to quenched disorder of the PDW SC state.

\subsection{Charge $4e$ SC order and the topological excitations of the PDW SC state}

If there is a uniform charge $4e$ SC order, its vortices must quantized in units of $hc/4e$ instead of the conventional $hc/2e$ flux quantum. Hence, half-
vortices are natural in this state. To see how they arise let us consider a system deep in the PDW SC state so that the magnitude of all the order 
parameters is essentially constant, but their phase may vary. Thus we can write the PDW SC order parameter as
\beq
\Delta(\vec{r})=|\Delta_{\vec{Q}}| \; e^{i \vec{Q} \cdot \vec{r}+i \theta_{\pm \vec{Q}}(\vec{r})}+
|\Delta_{-\vec{Q}}| \; e^{-i \vec{Q} \cdot \vec{r}+i \theta_{- \vec{Q}}(\vec{r})}
\eeq
where (by inversion symmetry) $|\Delta_{\vec{Q}}|=|\Delta_{- \vec{Q}}|=$const. It will be convenient to define the new phase fields $\theta_\pm
(\vec{r})$ by
\beq
\theta_{\pm \vec{Q}} (\vec{r})= \frac{1}{2} \left(\theta_+(\vec{r})\pm \theta_-(\vec{r})\right)
\eeq
Likewise, in the same regime the CDW order parameter can be written as
\beq
\rho(\vec{r})=|\rho_{\vec{K}}|\; \cos(\vec{K}\cdot \vec{r}+\phi(\vec{r}))
\eeq
(and a similar expression for the SDW order parameter.) In this notation, the second trilinear term shown in Eq.\eqref{eq:F3} takes the form
\beq
\mathcal{F}_{3,\gamma}=2 \gamma_\Delta |\rho_{\vec{K}} \Delta_{\vec{Q}} \Delta_{-\vec{Q}}| \; \cos(2 \theta_-(\vec{r})-\phi(\vec{r}))
\eeq
Hence, the {\em relative phase} $\theta_-$ is locked to $\phi$, the Goldstone boson of the CDW (the phason), and are not independently fluctuating fields. 
Furthermore, the phase fields $\theta_{\pm  \vec{Q}}$ are defined modulo $2\pi$ while $\theta_+$ is defined only modulo $\pi$.

This analysis implies that the allowed topological excitations of the PDW SC are
\begin{enumerate}
\item
A conventional SC vortex with  $\Delta \theta_+=2\pi$ and $\Delta \phi=0$, with topological charges $(1,0)$.
\item
A bound state of a 1/2 vortex and a CDW dislocation, 
$\Delta \theta_+=\pi$ and  $\Delta \phi=2\pi$, with topological charges $(\pm 1/2,\pm 1/2)$ (any such combination is allowed).
\item
A double dislocation, $\Delta \theta_+ =0$ and $\Delta \phi=4\pi$, with topological charge $(0,1)$.
\end{enumerate}
All three topological defects have logarithmic interactions.

There are now three different pathways to melt the PDW SC\cite{berg-2009}, depending which one of these topological excitations becomes relevant (in 
the Kosterlitz-Thouless RG sense\cite{kosterlitz-1973}) first. At temperature $T$, the scaling dimension of a topological excitation of topological charge $
(p,q)$ is
\beq
\Delta_{p,q}=\frac{\pi}{T} \left(\rho_{sc} p^2+\kappa_{CDW} q^2\right)
\eeq
where $\rho_{sc}$ is the superfluid density (the stiffness of the $\theta_+$ phase field) and $\kappa_{CDW}$ is the CDW stiffness (that is, of the $\phi$ 
phase field). As usual the criterion of relevance is that an operator that creates and excitation is relevant if its scaling dimension is equal to the space 
dimension (for details see, for instance, Ref.\cite{cardy-book}) which in this case is $2$. This condition, $\Delta_{p,q}=2$ for each one of the topological 
excitations listed above, leads to the phase thermal phase diagram shown in Fig.\ref{fig:pdw-phase-diagram}.\footnote{A more elaborate version of this 
phase diagram, based on a one-loop Kosterlitz RG calculation for for physically very different systems with the same RG structure system, was given in 
Refs.\cite{kruger-2002,podolsky-2007,radzihovsky-2008}.} 

Thus, the PDW state may thermally melt in three possible ways:
\begin{enumerate}
\item
First into a CDW phase (by proliferating conventional SC vortices, a $(1,0)$ topological excitation, followed by a subsequent melting of the CDW into the 
normal (Ising nematic) high temperature phase. This scenario corresponds to the right side of the phase diagram and , presumably, is what happens in 
{\LBCO}.
\item
A direct melting into the normal (Ising nematic) phase by proliferation of fractional vortices, with topological charge $(\pm 1/2, \pm 1/2)$.
\item
Melting into a charge $4e$ uniform SC phase by proliferation of double dislocations, with topological charge $(0,1)$.
\end{enumerate}

\begin{figure}
\begin{center}
\includegraphics[width=0.6\textwidth]{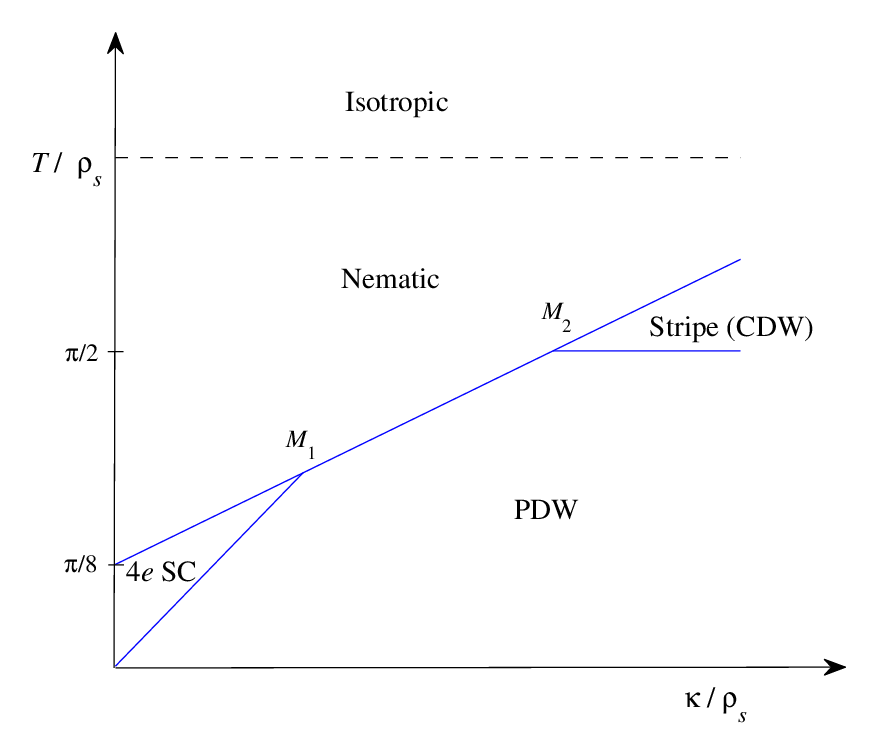}
\end{center}
\caption{Schematic phase diagram of the thermal melting of the PDW state.}
\label{fig:pdw-phase-diagram}
\end{figure}

The prediction that the PDW state should effectively have a uniform charge $4e$ SC order with an anomalous $hc/4e$ flux quantization leads to a direct 
test of this state. this can be done by searching for fractional vortices, and similarly of fractional periodicity in the Josephson effect (and Shapiro steps). 
Similarly, the prediction that in the phase in which an uniform (d-wave) SC is present there should be a charge-ordered state with period equal to that of 
the SC (and of the SDW) is another direct test of this theory.

\section{Nematic Phases in Fermi Systems} 
\label{sec:nematic}

We now turn to the theory of the nematic phases. The nematic phase is the simplest of the liquid crystal states. In this state the system is electronically 
uniform but anisotropic. There are two ways to access this phase. One is by a direct transition from the isotropic electronic fluid. The other is by melting 
(thermal or quantum mechanical) the stripe phase. We will consider both cases. We will begin with the first scenario in its simplest description as a {\em 
Pomeranchuk instability} of  a Fermi liquid. 

\subsection{The Pomeranchuk Instability}
\label{sec:pomeranchuk}

The central concept of the Landau theory of the Fermi liquid\cite{baym-1991} is the quasiparticle. A Landau quasiparticle is the elementary excitation of a 
Fermi liquid with the same quantum numbers as a non-interacting electron. A necessary condition for the Landau theory to work is the condition that the 
quasiparticle becomes sharp (or well defined) at asymptotically low energies, {\it i.e.\/} as the Fermi surface is approached. For the Landau quasiparticle 
to be well defined it is necessary that the quasiparticle width, {\it i.e.\/} the quasiparticle scattering rate, to be small on the scale of the quasiparticle energy. 
The quasiparticle scattering rate, the imaginary part of the electron self energy, $\Sigma^{\prime\prime}(\omega,{\vec{p}})$, is determined by the 
quasiparticle interactions, which in the Landau theory of the Fermi liquid are parametrized by the {\em Landau parameters}. Except for the BCS channel, 
the forward scattering  interactions (with or without spin flip) are the only residual interactions among the quasiparticles that survive at low 
energies.\cite{polchinski-1992,shankar-1994}

The Landau ``parameters'' are actually functions $F^{S,A}(\vec{p}, \vec{p}')$
quantify the strength of the forward scattering interactions among quasiparticles at low energies with momenta $\vec{p}$ and $\vec{p}'$ close to the Fermi 
surface in the singlet (charge) channel (S) or the triplet (spin) channel (A).   For  a translationally invariant system it depends only on the difference of the 
two momenta, $F(\vec{p}, \vec{p}')=F(\vec{p} - \vec{p}')$. Furthermore, if the system is also rotationally invariant, the Landau parameters can be expressed in 
an angular momentum basis. In 3D they take the form $F^{S,A}_{\ell,m}$ (with $\ell=0,1,2,\ldots$ and $ |m|\leq \ell$) , while in 2D they are simply 
$F^{S,A}_m$ (where $m \in \mathbb{Z}$). We will see below that in some cases of interest we will also need to keep the dependence on a small 
momentum transfer in the Landau parameters ({\it i.e.\/} $\vec{p}$ and $\vec{p}'$ will not be precisely at the FS) even though it amounts to keeping a 
technically irrelevant interaction. 
On the other hand, for a lattice model rotational invariance is always broken down to the point (or space) group symmetry of the lattice. In that case the 
Landau parameters are classified according to the irreducible representations of the point (or space) group of the lattice, {\it e.g.\/} the $\mathcal{C}_4$ 
group of  the square lattice.

It is well known in the Landau theory of the Fermi liquid that the thermodynamic stability of the Fermi liquid state requires that the Landau parameters 
cannot be too negative. This argument, due to Pomeranchuk\cite{pomeranchuk-1958}, implies that if in one channel the forward scattering interaction 
becomes sufficiently negative (attractive)  to overcome the stabilizing effects of the Pauli pressure, the Fermi liquid becomes unstable to a distortion of the 
FS with the symmetry of the unstable channel.\footnote{Although the Pomeranchuk argument is standard and reproduced in all the textbooks on Fermi 
liquid theory (see, {\it e.g.\/} Ref. \cite{baym-1991}) the consequences of this instability were not pursued until quite recently.}

Oganesyan {\it et al}\cite{oganesyan-2001} showed that in a 2D system of interacting fermions, the Pomeranchuk instability in fact marks a quantum 
phase transition to a {\em nematic Fermi fluid}. We will discuss this theory below in some detail. While the the theory of Oganesyan and coworkers 
applied to a system in the continuum, Kee and coworkers\cite{kee-2003,khavkine-2004} considered a lattice model. Hints of nematic order in specific 
models had in fact been discovered independently (but not recognized as such originally), notably by the work of Metzner and 
coworkers\cite{yamase-2005,halboth-2000,metzner-2003,neumayr-2003,yamase-2005,dellanna-2006}\footnote{In fact, perturbative renormalization 
group calculations\cite{honerkamp-2001,honerkamp-2002} have found a runaway flow in the $d_{x^2-y^2}$ particle-hole channel, which is a nematic 
instability, but it was not recognize it as such. See, however, Ref.\cite{hankevych-2002}.} 

There is by now a growing literature on the nematic instability. 
Typically the models, both in the continuum\cite{oganesyan-2001} or on different lattices\cite{kee-2003,lamas-2008a,lamas-2008a,quintanilla-2008b}, 
are solved within a 
Hartree-Fock type approximation (with all 
the limitations that such an approach has), or in special situations such as vicinity to Van Hove singularities\cite{khavkine-2004,yamase-2005} and certain 
degenerate band crossings\cite{sun-2009} (where the theory is better 
controlled), or using uncontrolled approximations to strong coupling systems such as slave fermion/boson 
methods.\cite{yamase-2000,miyanga-2006}. A strong coupling limit of the Emery model of the cuprates was shown to have a nematic state 
in Ref.\cite{kivelson-2004} (we will review this work below). Finally some non-perturbative work on the nematic quantum phase transition has been done 
using higher dimensional bosonization in Refs.\cite{lawler-2006,lawler-2007} and by RG methods.\cite{metlitski-2010}

Extensions of these ideas have been applied to the problem of the nematic phase seen in the metamagnetic 
bilayer ruthenate {\SROtwo} relying either on the van Hove mechanism\cite{kee-2005,yamase-2007c,puetter-2007,puetter-2009} 
or on an orbital ordering mechanism\cite{raghu-2009,lee-2009}, and in the new iron-based  superconducting compounds.\cite{fang-2008,xu-2008} More 
recently nematic phases of 
different types have been argued to occur in dipolar Fermi gases  of ultra-cold atoms.\cite{fregoso-2009,fregoso-2009b}\footnote{Another class of nematic 
state can occur inside a $d_{x^2-y^2}$ superconductor. This quantum phase transition involves primarily the nodal quasiparticles of the superconductor 
and it is tractable within large $N$ type approximations.\cite{kim-2008c,huh-2008}}

\subsection{The Nematic Fermi Fluid}
\label{sec:nematic-phase}

Here I will follow the work of Oganesyan, Kivelson and Fradkin\cite{oganesyan-2001} and consider first the instability in the charge (symmetric) channel. 
Oganesyan {\it et al} defined a charge nematic order parameter for two-dimensional Fermi fluid is the $2 \times 2$ symmetric traceless tensor of the form
\begin{equation}
\hat{\mathcal{Q}}(x)\equiv
-\frac{1}{k^2_F} {\Psi}^{\dagger}(\vec{r})
\left( 
\begin{array}{cc}
\partial^2_x-\partial^2_y & 2\partial_x\partial_y\\ 
2\partial_x\partial_y     & \partial^2_y-\partial^2_x
\end{array}
\right) \Psi(\vec{r}),
\label{eq:Q}
\end{equation}
It can also be represented by a complex valued field $\mathcal{Q}_2(x)$ whose expectation value is the nematic phase is
\begin{equation}
 \langle \mathcal{Q}_2\rangle \equiv \langle \Psi^\dagger \left(\partial_x + i \partial_y\right)^2\Psi\rangle=  |\mathcal{Q}_2|\; e^{2i\theta_2}=\mathcal{Q}_{11} 
 + i \mathcal{Q}_{12}\neq 0
 \label{eq:director}
 \end{equation}
$\mathcal{Q}_2$ transforms under rotations in the representation of angular momentum $\ell=2$.  
Oganesyan {\it et al} showed that if $ \langle \mathcal{Q}_2\rangle \neq 0$ then the Fermi surface {\em spontaneously} distorts and becomes an ellipse 
with eccentricity $\propto Q$. This state breaks rotational invariance mod $\pi$.

More complex forms of order can be considered by looking at particle-hole condensates with angular momenta $\ell >2$ (see Ref.\cite{sun-2008b})
\begin{equation}
\langle \mathcal{Q}_\ell \rangle=\langle \Psi^\dagger \left(\partial_x +i \partial_y\right)^\ell \Psi \rangle
\label{eq:director-ell}
\end{equation}
For $\ell$ {\em odd}, this condensate breaks rotational invariance (mod $2\pi/\ell$). It also breaks parity $\mathcal{P}$ and time reversal $\mathcal{T}$ but 
$\mathcal{P} \mathcal{T}$ is invariant. For example the condensate with $\ell =3$ is effectively equivalent to the 
``Varma loop state''.\cite{varma-1997,varma-2005}. The states with  $\ell$ {\em even} are also interesting, {\it e.g.\/} a hexatic state is described by a 
particle-hole condensate with $\ell=6$.\cite{barci-2003}

In a 3D system, the anisotropic state is described by an order parameter $\mathcal{Q}_{ij}$ which is a traceless symmetric tensor (as in conventional 
liquid crystals \cite{chaikin-1995,degennes-1993}). More generally, we can define an order parameter that transforms under the $(\ell,m)$ representation 
of the group of $SO(3)$ spatial rotations.

Oganesyan {\it et al} considered in detail  Fermi liquid type model of the nematic transition and developed a (Landau) theory of the transition (``Landau on 
Landau''). The Hamiltonian of this model describes (spinless) fermions in the continuum with a two-body interaction corresponding to the $\ell=2$ particle-
hole angular momentum channel. The Hamiltonian is
\begin{equation}
H=\int d\vec{r} \; {\Psi}^{\dagger}(\vec{r}) \epsilon(\vec{\nabla} )\Psi(\vec{r}) 
+ \frac{1}{4}\int d\vec{r} \int d\vec{r}' 
F_2(\vec{r}-\vec{r}'){\rm Tr}[\hat{\mathcal{Q}}(\vec{r})\hat{\mathcal{Q}}(\vec{r}')]
\label{eq:H-OKF}
\end{equation}
where the free-fermion dispersion (near the FS) is
$\epsilon(\vec{k})=v_{F}q[1+a(\frac{q}{k_{F}})^{2}]$ (here $q\equiv |\vec{k}|-k_F$), and the interaction is given in terms of the coupling 
\beq
F_2(\vec{r}) =(2\pi)^{-2}\int d\vec{q} e^{i\vec{q} \cdot \vec{r}}\frac{F_2}{1 +\kappa F_2 q^2}
\label{eq:F2q}
\eeq
where $F_2$ is the $\ell=2$  Landau parameter, and $\kappa$ measures the range of these interactions. Notice that we have kept a cubic momentum 
dependence in the dispersion, which is strongly irrelevant in the Landau Fermi liquid phase (but it is needed to insure stability in the broken symmetry 
state).

The Landau energy density functional for this model has the form (which can be derived by Hartree-Fock methods or, equivalently, using a Hubbard-
Stratonovich decoupling)
\beq
\mathcal{U}[{\mathcal{Q}}] =\mathcal{E}({\mathcal{Q}}) -\frac {\tilde \kappa} 4 {\rm Tr}[{\mathcal{Q}}
{\vec D}{\mathcal{Q}} ] - \frac {{\tilde \kappa}'} 4 {\rm Tr}[ {\mathcal{Q}}^2{\vec D}{\mathcal{Q}} ]+\ldots 
\label{eq:Landau-U}
\eeq
where $\kappa$ and $\tilde \kappa$ are the two effective Franck constants (see Ref.\cite{chaikin-1995}). The uniform part of the energy functional, 
$\mathcal{E}(\mathcal{Q})$, is given by
\beq
\mathcal{E}({\mathcal{Q}})=\mathcal{E}({\vec 0})+\frac{A}{4}{\rm Tr}[{\mathcal{Q}}^2]+ 
\frac{B}{8}{\rm Tr}[{\mathcal{Q}}^4]+\ldots
\label{eq:E(Q)}
\eeq
where 
\beq
A=\frac{1}{2N_F}+F_2
\label{eq:A}
\eeq
$N_F$ is the density of states at the Fermi 
surface, and the coefficient of the quartic term is 
\beq
B=\frac{3aN_F|F_2|^3}{8 E^2_F}
\label{eq:B}
\eeq
$E_F\equiv v_Fk_F$ is the Fermi energy.\cite{oganesyan-2001}
The (normal) Landau Fermi liquid phase is stable provided $A>0$, or, what is the same, if $2N_FF_2> -1$ which is the Pomeranchuk condition (in this 
notation). On the other hand, thermodynamic stability also requires that $B>0$, which implies that the coefficient of the cubic correction in the dispersion 
be positive, $a>0$. If this condition is not satisfied, as it is the case in simple lattice models\cite{kee-2003}, higher order terms must be kept to insure 
stability. However, in this case the transition is typically first order.

This model has two phases: 
\begin{itemize}
\item
an isotropic Fermi liquid phase, $A>0$
\item
a nematic (non-Fermi liquid) phase, $A<0$ 
\end{itemize}
separated by a quantum critical point at the Pomeranchuk value, $2N_F F_2=-1$.

Let us discuss the quantum critical behavior. We
will parametrize the distance to the Pomeranchuk QCP by 
\beq
\delta=\frac{1}{2N_F}+ F_2
\label{eq:delta}
\eeq
and define $s=\omega/qv_F$.
The transverse collective nematic modes have Landau damping at the QCP.\cite{oganesyan-2001} Their effective action has the form
\beq
S_\perp=\int d\omega d {\vec{q}} \; \big(\kappa q^2+\delta-i \frac{|\omega|}{qv_F}\big) \; |\mathcal{Q}_\perp(\omega,\vec{q})|^2
\label{eq:Sperp}
\eeq
which implies that the dynamic critical exponent is $z=3$.\footnote{There are other collective modes at higher energies. In particular there is an 
underdamped longitudinal collective mode with $z=2$.\cite{oganesyan-2001} These higher energy modes contribute to various crossover 
effects\cite{zacharias-2009}, but decouple in the asymptotic quantum critical regime.} 

According to the standard perturbative criterion of 
Hertz \cite{hertz-1976} and Millis \cite{millis-1993}, the quantum critical behavior is that of an equivalent $\phi^4$ type field theory in dimensions $D=d+z$ 
which in this case is $D=5$. Since the upper critical (total) dimension is $4$, the Hertz-Millis analysis would predict that mean field theory is 
asymptotically exact and that the quartic (and higher) powers of the order parameter field are {\em irrelevant} at the quantum critical point (for an extensive 
discussion see Ref. \cite{sachdev-1999}.) However we will see below that while this analysis is correct for the bosonic sector of the theory, {\it i.e.\/} the 
behavior of the bosonic collective modes such as the order parameter itself, the situation is far less clear in the fermionic sector. We will come back to this 
question below.

Let us discuss now the physics of the nematic phase. In the nematic phase the FS is spontaneously distorted along the direction of the (director) order 
parameter (see Fig.\ref{fig:Pomeranchuk2}) and exhibits a quadrupolar ($d$-wave) pattern, {\it i.e.\/} the Fermi wave vector has an angular dependence 
$k_F(\theta) \propto \cos 2\theta$ (in 2D). Indeed, in the nematic phase the Hartree-Fock wave function 
is a Slater determinant whose variational parameters determine the shape of the FS.
\begin{figure}[h!]
\begin{center}
\includegraphics[width=0.5\textwidth]{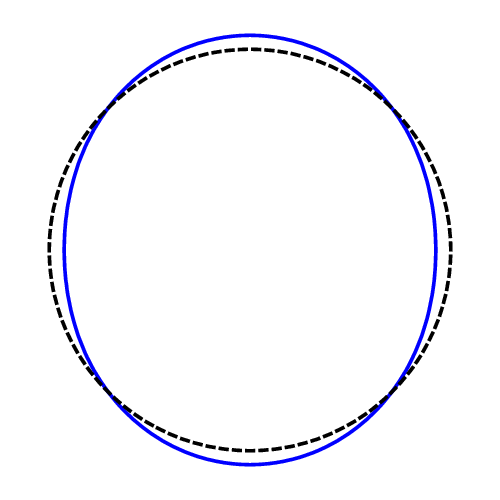}
\end{center}
\caption{Spontaneous distortion of the Fermi surface in the nematic phase of a 2D Fermi fluid.}
\label{fig:Pomeranchuk2}
\end{figure}

In principle, a wave function with a similar structure can be used to suggest (as it was done in Ref.\cite{oganesyan-2001}) that it should also apply to the 
theory of the electronic nematic state observed in the 2DEG in large magnetic fields. In that framework one thinks of the 2DEG in a half-filled 
landau level  as an equivalent system of  ``composite fermions''\cite{jain-1989}, fermions coupled to a 
Chern-Simons gauge field\cite{lopez-1991,halperin-1993}. It has been argued\cite{rezayi-1994} that this state can be well described by a Slater 
determinant wave function, projected onto the Landau level. The same procedure can be applied to the nematic wave function, and some work has been 
done along this direction\cite{manousakis-2008}. A problem that needs to be solved first is the determination  of the Landau parameters of the composite 
fermions of which very little (that makes sense) is known.

A simple (Drude) calculation then shows that the transport is anisotropic. The resistivity in the nematic phase,  due to scattering from structureless 
isotropic impurities, yields the result that the resistivity tensor is anisotropic with an anisotropy controlled by the strength of the nematic order parameter:
\beq
\frac{\rho_{xx}-\rho_{yy}}{\rho_{xx}+\rho_{yy}}=\frac{1}{2}\frac{m_y-
m_x}{m_y+m_x}=\frac{{\rm Re} \; Q}{E_F}  + O(Q^3)
\label{eq:transport-anisotropy}
\eeq
where $m_x$ and $m_y$ are the (anisotropic) effective masses of the quasiparticles in the nematic state. In general it is a more complex odd function of 
the order parameter.

In the nematic phase the transverse Goldstone boson is generically {\em overdamped} (Landau 
damping) except for a finite set of symmetry directions, $\phi=0,\pm \pi/4,\pm \pi/2$, where it is {\em underdamped}. Thus, $z=3$ scaling also applies to 
the nematic phase for general directions. Naturally, in a lattice system the rotational symmetry is not continuous and the transverse Goldstone modes are 
gapped. However, the continuum prediction still applies if the lattice symmetry breaking is weak and if either th energy or the temperature is larger that the 
lattice anisotropy scale.

On the other hand, the behavior of the fermionic correlators is much more strongly affected. To one loop order, the
quasiparticle scattering rate, $\Sigma^{''}(\omega,\vec{p})$ is found to have the behavior
\beq
\Sigma''(\omega,\vec{k}) =\frac{\pi}{\sqrt{3}}\frac{(\kappa k^2_F)^{1/3}}{\kappa N_F}
\left | \frac {k_{x}k_{y}} {k_F^2} \right|^{4/3}
\left |\frac{\omega}{2 v_F k_F}\right|^{2/3} +
\ldots
\label{eq:nematic-nfl}
\eeq
for $\vec{k}$ along a general direction. 
On the other hand, along a symmetry direction
\beq
\Sigma''(\omega)=
\frac{\pi}{3 N_{F} \kappa} \frac{1}{(\kappa k^2_F)^{1/4}}\left | \frac{\omega}{v_F k_F} \right |^{3/2}+\ldots
\label{eq:symmetry}
\eeq
Hence, the entire nematic phase is a non-Fermi liquid (again, with the caveat on lattice symmetry breaking effects). 

At the Pomeranchuk quantum critical point  the quasiparticle scattering rate obeys the same (one loop) scaling shown in Eq.\eqref{eq:nematic-nfl}, 
$\Sigma^{''}(\omega) \propto |\omega|^{2/3}$, both in continuum\cite{oganesyan-2001} and lattice models\cite{dellanna-2006}, but it is isotropic. In the 
quantum critical regime the electrical resistivity obeys a $T^{4/3}$ law.\cite{dellanna-2007}. Also, both in the nematic phase (without lattice anisotropy) 
and in the quantum critical regime, the strong nematic fluctuations yield an electronic contribution to the specific heat that scales as $T^{2/3}$ (consistent 
with the general scaling form $T^{d/z}$\cite{sachdev-1999}) which dominates over the standard Fermi liquid linear $T$ dependence at low 
temperatures.\cite{baym-1991} 

Since $\Sigma''(\omega) \gg \Sigma'(\omega)$ (as $\omega \to 0$), we need to asses the validity of these results as they signal a {\em failure of 
perturbation theory}. To this end we have used higher dimensional bosonization as a non-perturbative 
tool\cite{haldane-1994,CastroNeto-1993,CastroNeto-1995,houghton-1993,houghton-2000}. Higher dimensional bosonization reproduces the collective 
modes found in Hartree-Fock+ RPA and is consistent with the Hertz-Millis analysis of quantum criticality: 
$d_{\rm eff}=d+z=5$. \cite{lawler-2006,lawler-2007}. Within this approach, the fermion propagator takes the form
\beq
G_F(x,t)=G_0(x,t) Z(x,t)
\label{eq:GF}
\eeq
At the Nematic-FL QCP it has the form
\beq
G_F(x,0)=G_0(x,0)\; e^{\displaystyle{- {\rm const}.\, |x|^{1/3}}}
\label{eq:equal-time}
\eeq
at equal times, and 
\beq
G_F(0,t)=G_0(0,t)\; e^{\displaystyle{- {\rm const}.\, |t|^{-2/3}\ln t}}
\label{eq:equal-position}
\eeq
at equal positions. Notice that these expressions are consistent with the expected $z=3$ scaling even though the time and space dependence 
is not a power law.
The quasiparticle residue is then seen to vanish at the QCP:
\beq
Z=\lim_{x \to \infty} Z(x,0)=0
\label{eq:Z-qcp}
\eeq
However, the single particle density of states, $N(\omega)=-\frac{1}{\pi} \textrm{Im} G(\omega,0)$, turns out to have a milder behavior:
\beq
N(\omega)=N(0)\left(1-{\rm const}'. |\omega|^{2/3} \ln \omega\right)
\label{eq:DOS}
\eeq

\begin{figure}
\psfrag{EF}{$k_F$}
\psfrag{Z}{$Z$}
\psfrag{E(k)}{$k$}
\psfrag{n(k)}{$n(k)$}
\begin{center}
\includegraphics[width=0.5\textwidth]{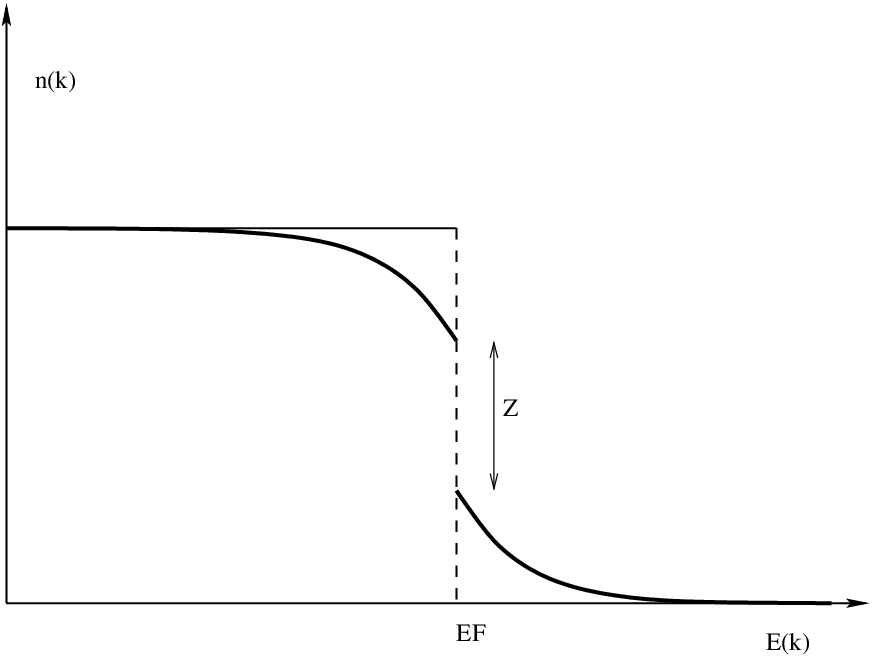}
\end{center}
\end{figure}

Let us now turn to the behavior near the QCP. For $T=0$ and $\delta\ll 1$ (on the Fermi Liquid side) the quasiparticle residue is now finite
\beq
Z\propto e^{\displaystyle{-{\rm const}./ \sqrt{\delta}}}
\label{eq:Z-Landau}
\eeq
but its dependence on the distance to the nematic QCP is an essential singularity. On the other hand,
right at the QCP $(\delta=0)$, and for a temperature range $T_F \gg T \gg T_\kappa$, the equal-time fermion propagator is found to vanish exactly 
\beq
Z(x,0)\propto e^{\displaystyle{-{\rm const}.\, T x^2 \ln \left(L/x\right)}}\to 0 \qquad \textrm{as} \; L \to \infty
\label{eq:Z-equal-time}
\eeq
but, the equal-position propagator $Z(0,t)$ remains {\em finite} in the thermodynamic limit, $L \to \infty$! This behavior has been dubbed 
``Local quantum criticality''.\footnote{A similar behavior was found in the quantum Lifshitz model at its QCP\cite{ghaemi-2005}.} On the other hand, 
irrelevant quartic interactions of strength $u$ lead to a renormalization of $\delta$ that smears the QCP at 
$T>0$
\cite{millis-1993}
\beq
\delta \to \delta(T)=-u T \ln \left(uT^{1/3}\right)
\label{eq:delta-renormalization}
\eeq
leading to a milder behavior at equal-times
\beq
Z(x,0)\propto e^{\displaystyle{-{\rm const}.\, T x^2 \ln(\xi/x)}} \qquad {\rm where}\;\;  \xi=\delta(T)^{-1/2}
\label{eq:Z-equal-time-renormalized}
\eeq

These results are far from being universally accepted. Indeed Chubukov and coworkers\cite{chubukov-2004b,chubukov-2005a,rech-2006} have argued 
that the perturbative non-Fermi liquid behavior, $\Sigma^{''}(\omega) \sim \omega^{2/3}$, which is also found at a ferromagnetic metallic QCP,  persists to 
all orders in perturbation theory and can recover the results of higher dimensional bosonization only by taking into account the most infrared divergent 
diagrams. The same non-Fermi liquid one-loop perturbative scaling has been found in other QCPs such as in the problem of fermions (relativistic or not) 
at finite density coupled to dynamical gauge fields. This problem has been discussed in various settings ranging from hot and dense QED and 
QCD\cite{holstein-1973,baym-1990,boyanovsky-2001}, to the gauge-spinon system in RVB approaches to high $T_c$ 
superconductors\cite{reizer-1989,ioffe-1990,nagaosa-1991,polchinski-1994,chakravarty-1995,ss_lee-2009} to the compressible states of the 2DEG in 
large magnetic fields\cite{halperin-1993}. In all cases these authors have also argued that the one-loop scaling persists to all orders. In a recent paper 
Metlitski and Sachdev \cite{metlitski-2010}  found a different  scaling behavior. 

We end with a brief discussion on the results in lattice models of the nematic quantum phase transition. This is important since, with the possible 
exception  of the 2DEG in large magnetic fields and in ultra-cold atomic systems, all strongly correlated systems of interest have very strong lattice effects. 
The main difference between the results in lattice models and in the continuum is that in the former the quantum phase transition (at the mean field, 
Hartree-Fock, level) has a strong tendency to be first order. Although fluctuations can soften the quantum transition and turn the system quantum critical 
(as emphasized in Ref.\cite{jakubczyk-2009}), nevertheless there are good reasons for the transition to be first order more or less generically. One is that if 
the stabilizing quartic terms are negative ({\it e.g.\/} say due to the band structure), this also results, in the case of a lattice system, in a  Lifshitz transition at 
which the topology of the FS changes from closed to open. This cannot happen in a continuous way.

\subsection{Generalizations: Unconventional Magnetism and Time Reversal Symmetry Breaking}
\label{sec:generalizations}

We will now consider briefly the extension of these ideas to the spin-triplet channel\cite{wu-2004,wu-2007}. In addition to particle-hole condensates in the 
singlet (charge) channel we will be interested particle-hole condensates in the spin (triplet) channel. In 2D the order parameters for particle-hole 
condensates in the spin triplet channel are (here $\alpha,\beta=\uparrow,\downarrow$)
\beq
\mathcal{Q}^a_\ell(r)=\langle\Psi_\alpha^\dagger(r) \sigma^a_{\alpha \beta} \left(\partial_x+i\partial_y\right)^\ell\Psi_\beta(r)\rangle \equiv n_1^a+i n^a_2
\label{eq:Q-triplet}
\eeq
These order parameters transform under both $SO(2)$ spatial rotations and under the internal $SU(2)$ symmetry of spin. If
$\ell\neq 0$ the state has a broken rotational invariance in space and in spin space. These states are a particle-hole condensate analog of the 
unconventional superconductors and superfluids, such as He$_3$A and He$_3$B. Indeed one may call these states ``unconventional magnetism'' as the 
$\ell=0$ (isotropic) state is just a ferromagnet. In 2D these states are then given in terms of two order parameters, each in the vector (adjoint) 
representation of the $SU(2)$ spin symmetry.\footnote{In 3D the situation is more complex and the possible are more subtle. In particular, in 3D there are 
three vector order parameters involved.\cite{wu-2007}} We will discuss only the 2D case.
The order parameters obey the following transformation laws:
\begin{enumerate}
\item
Time Reversal: 
\beq
\mathcal{T} \mathcal{Q}^a_\ell \mathcal{T}^{-1}=(-1)^{\ell+1} \mathcal{Q}^a_\ell
\label{eq:time-reversal}
\eeq
\item
Parity: 
\beq
\mathcal{P} \mathcal{Q}^a_\ell \mathcal{P}^{-1}=(-1)^\ell \mathcal{Q}^a_\ell
\label{eq:parity}
\eeq
\item
$\mathcal{Q}^a_\ell$ rotates under an $SO_{\rm spin}(3)$ transformation, and transforms as $\mathcal{Q}^a_\ell \to \mathcal{Q}^a_\ell e^{i \ell \theta}$ 
under a rotation in space by an angle $\theta$.
\item
$\mathcal{Q}^a_\ell$ is invariant under a rotation by $\pi/\ell$ followed by a spin flip.
\end{enumerate}

Wu and collaborators\cite{wu-2004,wu-2007} have shown that these phases can also be accessed by a Pomeranchuk instability in the spin (triplet) 
channel.\footnote{The $\ell=0$ case is, of course, just the conventional Stoner ferromagnetic instability.} They showed that the Landau-Ginzburg free 
energy takes the simple form
\beq
F[n]=r (|\vec{n}_1|^2+|\vec{n}_2|^2)+ v_1 (|\vec{n}_1|^2+|\vec{n}_2|^2)^2 + v_2 |\vec{n}_1 \times \vec{n}_2|^2
\label{eq:FLG-triplet}
\eeq
The Pomeranchuk instability occurs at $r=0$, {\it i.e.} for  $N_FF_\ell^A=-2$ (with $ \ell \geq 1)$, where $F^A_\ell$ are the Landau parameters in the spin-
triplet channel. Notice that this free energy is invariant only by global $SO(3)$ rotations involving both vector order parameters, $\vec{n}_1$ and 
$\vec{n}_2$. Although at this level the $SO(3)$ invariance is seemingly an internal symmetry, there are gradient terms that lock the internal $SO(3)$ spin 
rotations to the ``orbital'' spatial rotations. (See Ref.\cite{wu-2007}). A similar situation also occurs in classical liquid crystals.\cite{chaikin-1995}

At the level of the Landau-Ginzburg theory the system has two phases with broken $SO(3)$ invariance:
\begin{enumerate}
\item
If $v_2>0$, then the two $SO(3)$ spin vector order parameters must be parallel to each other, $\vec{n}_1 \times \vec{n}_2=0$. They dubbed this the ``$
\alpha$'' phase. In the $\alpha$ phases the up and down Fermi surfaces are distorted (with a pattern determined by $\ell$) but are rotated from each other 
by $\pi/\ell$. One case of special interest is the $\alpha$ phase with $\ell=2$. This is the ``nematic-spin-nematic'' discussed briefly in 
Ref.\cite{kivelson-2003}.\footnote{The term ``nematic-spin-
nematic'' is a poor terminology. A spin nematic is a state with a magnetic order parameter that is a traceless symmetric tensor, which this state does not.}  
In this  phase the spin up and spin down FS have an $\ell=2$ quadrupolar (nematic) distortion but are rotated by $\pi/2$ (see Fig.\ref{fig:alpha}).
\begin{figure}[t!]
\psfrag{l=1}{$l=1$}
\psfrag{l=2}{$l=2$}
\psfrag{s}{$\vec{s}$}
\psfrag{a-phase}{\small $\alpha$-phase}
\psfrag{dk1}{$\delta k_\uparrow$}
\psfrag{dk2}{$\delta k_\downarrow$}
\begin{center}
\includegraphics[width=0.6\textwidth]{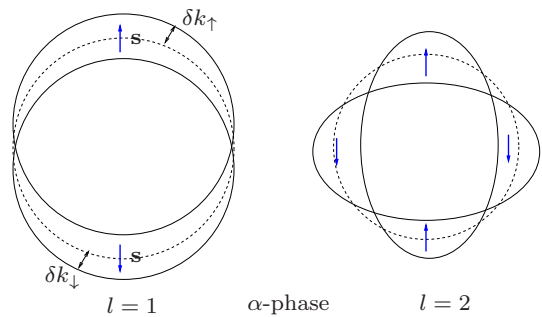}
\end{center}
\caption{The $\alpha$-phases in the $\ell=1$ and $\ell=2$ spin triplet channels. 
The Fermi surfaces exhibit the $p$ and $d$-wave distortions, respectively.}
\label{fig:alpha}
\end{figure}  
\item
Conversely, if $v_2<0$, then the two $SO(3)$ spin vector order parameters must be orthogonal to each other, $ \vec{n}_1 \cdot \vec{n}_2=0$ and $|\vec{n}_1|=|\vec{n}_2|$. 
\begin{figure}[t!]
\psfrag{s}{$\vec{s}$}
\psfrag{dk1}{$\delta k_\uparrow$}
\psfrag{dk2}{$\delta k_\downarrow$}
\begin{center}
\subfigure[~$w=2$]{\includegraphics[width=0.3\linewidth]{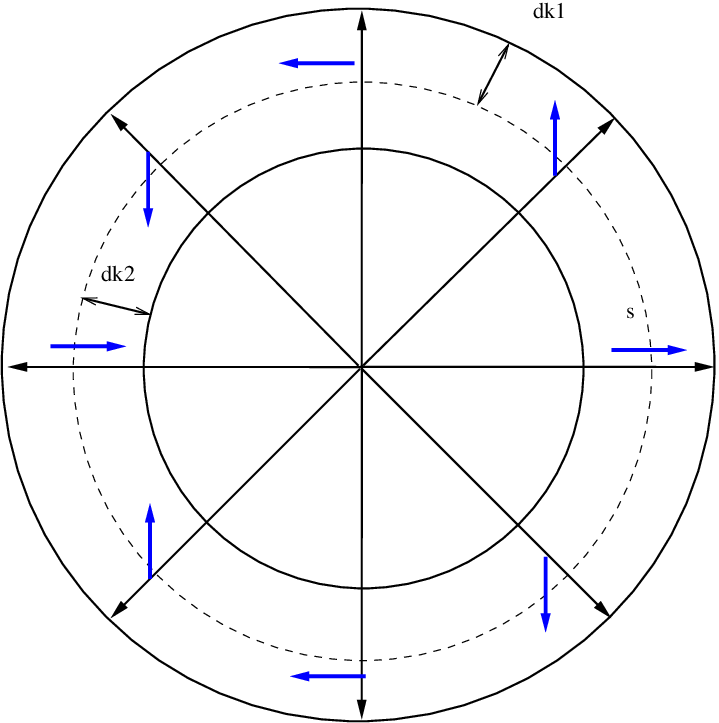}}
\subfigure[~$w=-2$]{\includegraphics[width=0.3\linewidth]{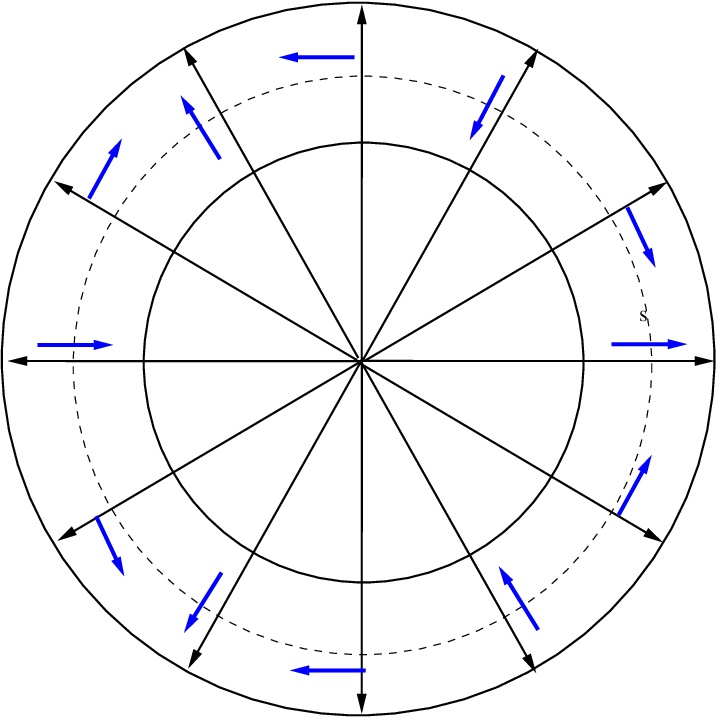}}
\end{center}
\caption{The $\beta$-phases in the $\ell=2$ triplet channel. Spin configurations
exhibit the vortex structure with winding number $w=\pm 2$.
These two configurations can be transformed to each other
by performing a rotation around the $x$-axis with the angle of $\pi$.}
\label{fig:vortex2}
\end{figure}   
Wu {\it et al} dubbed these the ``$\beta$'' phases. In the $\beta$ phases there are two isotropic FS but spin is not a good quantum 
number. In  fact, the electron self energy in the $\beta$-phases acquires a spin-orbit type form with a strength given by the magnitude of the order 
parameter. The  mean-field electronic structure thus resembles that of a system with a strong and tunable spin-orbit coupling ({\it i.e.\/} not of 
$O((v_F/c)^2)$ as it is normally the case.) 

We can define now a $\vec{d}$ vector: 
\beq
{\vec{d}}(\vec{k})=(\cos (\ell \theta_{\vec{k}}),\sin(\ell\theta_{\vec{k}}),0)
\label{eq:d-vector}
\eeq
 In the $\beta$ phases the $\vec{d}$ vector winds about the undistorted FS. For the special case of $\ell=1$, the windings are $w=1$ (corresponding to a 
 ``Rashba'' type state) and  $w=-1$ (corresponding to a ``Dresselhaus'' type state). For the $d$-wave case, the winding numbers are $w=\pm 2$ (see Fig.
 \ref{fig:vortex2}.)
\end{enumerate}
These phases have a rich phenomenology of collective modes and topological excitations which we will not elaborate on here. See Ref.\cite{wu-2007} for 
a detailed discussion.\footnote{The $p$-wave ($\ell=1$)  $\beta$ phase has the same physics as the `spin-split' metal of Ref.\cite{hirsch-1990}. A similar 
state was proposed in Ref.\cite{varma-2006} as an explanation of the ``hidden order'' phase of URu$_2$Si$_2$.}

Fermionic systems with dipolar magnetic interactions may be a good candidate for phases similar to the ones we just described 
(See Refs.\cite{fregoso-2009,fregoso-2009b}, and it is quite possible that these systems may be realized in ultra-cold atomic gases. In that context the 
anisotropic form of the dipolar interaction provides for a mechanism to access some of this physics. Indeed, in the case of a fully polarized (3D) dipolar 
Fermi gas, the FS will have an uniaxial distortion. If the polarization is spontaneous (in the absence of a polarizing external field) this phase is actually a 
ferro-nematic state, a state with coexisting ferromagnetism and nematic order. If the system is partially polarized then the phase is a mix of nematic order 
and  ferromagnetism coexisting with a phase with a non-trivial ``spin texture'' in momentum space.

It turns out that generalizations of the Pomeranchuk picture of the nematic state to multi-band electronic systems can describe metallic states with a 
spontaneous breaking of time reversal invariance. This was done in Ref.\cite{sun-2008b} where it was shown that in a two-band system ({\it i.e.\/} a 
system with two Fermi surfaces)  it is possible to have a metallic state which breaks time reversal invariance and exhibits a spontaneous anomalous Hall 
effect. While the treatment of this problem has a superficial formal similarity with the triplet (spin) case, {\it i.e.\/} regarding the band index as a ``pseudo-
spin'' (or flavor), the physics differs considerably. At the free fermion level the fermion number on each band is separately conserved, leading to a formal 
$SU(2)$ symmetry. However, the interacting system has either a smaller $U(1) \times U(1)$ invariance or, more generally, 
$\mathbb{Z}_2 \times \mathbb{Z}_2$ invariance, as the more general interactions preserve only the {\em parity} of the band fermion 
number.\cite{sun-2008b} At any rate it turns out that analogs of the ``$\alpha$'' and ``$\beta$'' phases of the triplet channel exist in multi-band systems. 
The ``$\alpha$'' phases break time reversal and parity (but not the product). An example of such metallic (gapless) states is the 
``Varma loop state''.\cite{varma-2005,simon-2002}. The ``$\beta$'' states break time reversal invariance (and chirality). In the ``$\beta$'' phases there is a 
spontaneous anomalous Hall effect, {\it i.e.\/} a zero field Hall effect with a Hall conductivity that is not quantized as this state is a metal\footnote{This is 
consistent with the general arguments of Ref.\cite{haldane-2004}.}, whereas the ``$\alpha$'' phases there is not.

\section{Nematic Order in the Strong Correlation Regime}
\label{sec:strong-coupling}

\begin{figure}
\psfrag{ud}{$U_d$}
\psfrag{up}{$U_p$}
\psfrag{eps}{$\epsilon$}
\psfrag{vpp}{$V_{pp}$}
\psfrag{vpd}{$V_{pd}$}
\psfrag{tpp}{$t_{pp}$}
\psfrag{tpd}{$t_{pd}$}
\begin{center}
\includegraphics[width=0.5\textwidth]{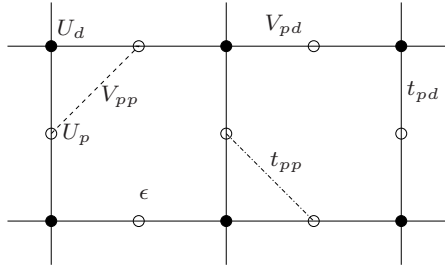}
\end{center}
\caption{The Emery model of the CuO lattice.}
\label{fig:emery}
\end{figure}

We will now discuss how a nematic state arises as the exact ground state in the strong coupling limit of the Emery model.\cite{kivelson-2004} The Emery 
model is a simplified microscopic model of the important electronic degrees of freedom of the copper oxides.\cite{emery-1987}  In this model, the CuO 
plane is described as a square lattice with the Cu atoms residing on the sites and the O atoms on the links (the medial lattice of the square lattice). On 
each site of the square lattice there is a single $d_{x^2-y^2}$ Cu orbital, and a $p_x$ ($p_y$) O orbital on each site of the medial along the $x$ ($y$) 
direction. We will denote by $d_\sigma^\dagger(\vec{r})$ the operator the creates a hole on the Cu site $\vec{r}$ and by 
$p^\dagger_{x,\sigma}(\vec{r}+\frac{\vec{e}_x}{2})$ and $p^\dagger_{y,\sigma}(\vec{r}+\frac{\vec{e}_y}{2})$ the operators the create a hole on the 
O  sites $\vec{r} +\frac{\vec{e}_x}{2}$  and $\vec{r} +\frac{\vec{e}_y}{2}$ respectively. 

The Hamiltonian of the Emery model is the sum of kinetic energy and  interaction terms. The kinetic energy terms consist of the hopping of a hole from a 
Cu site to its nearest O sites (with amplitude $t_{pd}$), an on-site energy 
$\varepsilon>0$ on the O sites (accounting for the difference in ``affinity'' between Cu and O), and a (small) direct hopping between nearest-neighboring O 
sites, $t_{pp}$. The interaction terms are just the on-site Hubbard repulsion $U_d$ (on the Cu sites) and $U_p$ (on the O sites) as well as nearest 
neighbor (``Coulomb'') repulsive interactions of strength $V_{pd}$ (between Cu and O) and $V_{pp}$ (between two nearest O). (See Fig.\ref{fig:emery}.)  It 
is commonly believed that this model is equivalent to its simpler cousin, the one band Hubbard model. However, while this equivalency is approximately 
correct in the weak coupling limit, it is known to fail already at moderate couplings. We will see that in the strong coupling limit, no such reduction (to a 
``Zhang-Rice singlet'') is possible.

Let us look at the energetics of the 2D Cu O model in the strong coupling limit. By strong coupling we will mean the regime in which the following 
inequalities hold:
\beq 
\frac{t_{pd}}{U_p}, \frac{t_{pd}}{U_d},  \frac{t_{pd}}{V_{pd}}, \frac{t_{pd}}{V_{pp}} \to 0, \qquad
U_d > U_p\gg V_{pd}>V_{pp}, \qquad \textrm{and} \qquad \frac{t_{pp}}{t_{pd}}\to 0
\eeq
as a function of hole doping $x>0$, where $x$ is the number of doped holes per Cu.
In this regime, neither Cu nor O sites can be doubly occupied. At half filling, $x=0$, the holes occupy all the Cu sites and all O sites are empty.
 At half-filling and in this strong coupling regime the Emery model (much as the Hubbard 
model) is equivalent to a quantum Heisenberg antiferromagnet with a small exchange coupling $J_H \approx \frac{8 t_{pd}^4}{U_pV_{pd}^2}$.
This is the double-exchange mechanism. (It turns out that in this model the four-spin ring exchange interactions can be of the same order 
of magnitude as  the Heisenberg exchange $J_H$.\cite{kivelson-2004}.)

Let us consider now the very low doping regime, $ x \to 0$. Any  additional hole will have to be on an O site. The 
energy to add one hole ({\it i.e.\/} the chemical potential $\mu$ of a hole) is $\mu\equiv 2V_{pd}+\epsilon$. Similarly, the
energy of two holes on nearby $O$ sites is $2\mu+V_{pp}$. It turns out that in this strong coupling regime, with $t_{pp}=0$, the dynamic of the doped 
holes is strongly constrained and effectively becomes one-dimensional.
The simplest allowed move for a hole on an O site, which takes two steps, is shown in Fig.\ref{fig:allowed-move}.
\begin{figure}[h!]
\begin{center}
\includegraphics[width=0.6\textwidth]{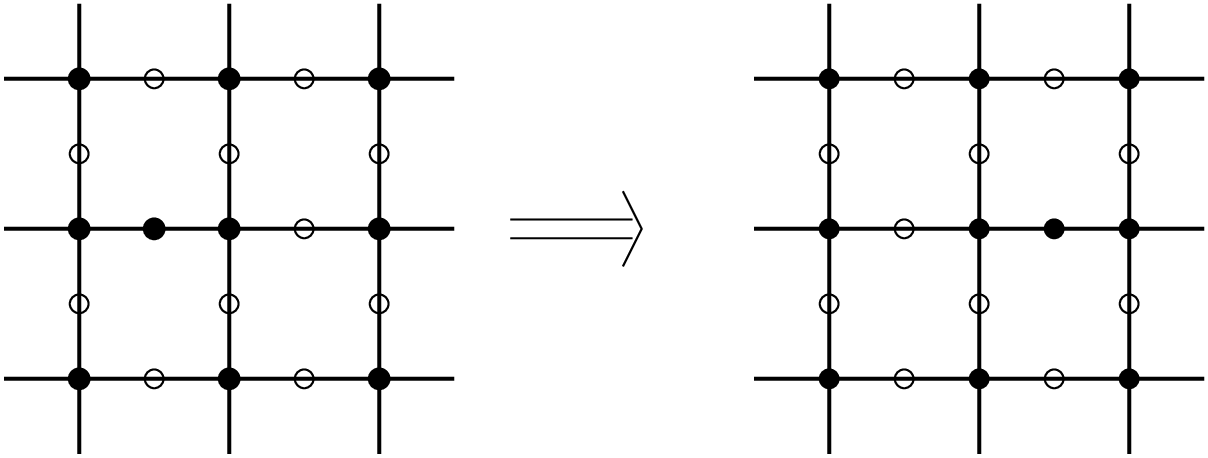} 
\end{center}
\caption{An allowed two-step move.}
\label{fig:allowed-move}
\end{figure}
The final and initial states are degenerate, and their energy is $E_0+\mu$, where $E_0$ is the ground state energy of the undoped system. If this was the 
only allowed process, the system would behave as a collection of 1D fermionic systems. 

To assess if this is correct let us examine other processes to the 
same (lowest) order in perturbation theory (in powers of the kinetic energy). One possibility is a process in which in the final state the hole ``turned the 
corner'' (went from being on an x oxygen to a near y oxygen). However for that to happen it will have to go through an intermediate state such as the one 
shown in Fig.\ref{fig:intermediate}a. This intermediate state has an energy $E_0+\mu+V_{pp}$. Hence, the effective hopping matrix element to turn the 
corner is $t_{\rm eff}=\frac{t_{pd}^2}{V_{pp}}\ll t_{pd}$, which is strongly suppressed by the Coulomb effects of $V_{pp}$. In contrast, the intermediate state 
for the hole to continue on the same row (see Fig.\ref{fig:intermediate}) is $E_0+\mu+\epsilon$. Thus the effective hopping amplitude instead becomes
$t_{\rm eff}=\frac{t_{pd}^2}{\epsilon}$, which is not suppressed by Coulomb effects of $V_{pp}$. All sorts of other processes have large energy 
denominators and are similarly suppressed (for a detailed analysis see Ref.\cite{kivelson-2004}.)
\begin{figure}[h!]
\begin{center}
\includegraphics[width=0.6\textwidth]{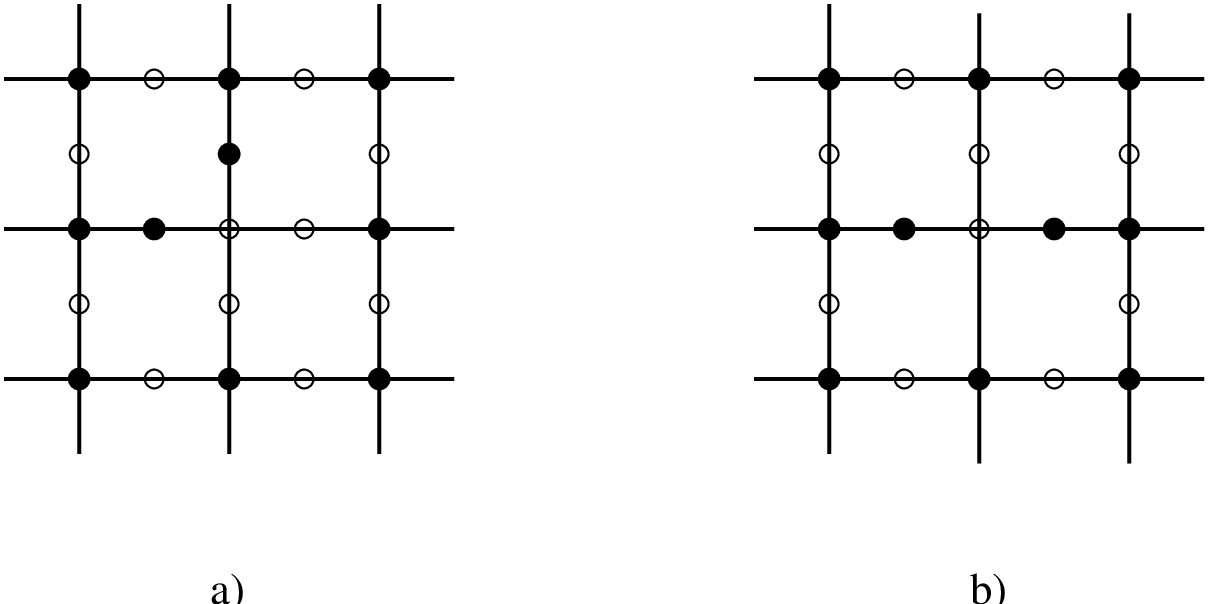} 
\end{center}
\caption{Intermediate states for processes in which a hole a) turns a corner and b) continues on the same row.}
\label{fig:intermediate}
\end{figure}
The upshot of this analysis is that in the strong coupling limit the oped holes behave like a set of one-dimensional fermions (one per row and column). 
The argument for one-dimensional effective dynamics can in fact be made more rigorous. In Ref.\cite{kivelson-2004} it is shown that to leading order in 
the strong coupling expansion  the system is is a generalized $t-J$ type model (with $J$ effectively set to zero). In this limit there are an infinite number of 
conserved charges, exactly one per row and one per column.

The next step is to inquire if, at fixed but very small doping $x \to 0$, the rows and columns are equally populated or not. Consider then two cases; a) all 
rows and columns have the same fermion density, and b) the columns (or the rows) are empty. Case a) is isotropic while case b) is nematic. It turns out 
that due to the effects of the repulsive Coulomb interaction $V_{pp}$, the nematic configuration has lower energy at low enough doping. The argument is 
as follows. For the nematic state (in which all rows are equally populated but all rows are empty), the ground state energy has an expansion in powers of 
the doping $x$ of the form:
\beq
E_{\rm nematic}=E(x=0)+\Delta_c \; x + W \; x^3 + O(x^5)
\label{eq:Enematic-emery}
\eeq
where $\Delta_c=2V_{pd}+\epsilon+\ldots$ and $W=\pi^2 \hbar^2/6m^*$. The energy of the isotropic state (at the same doping $x$) is
\beq
E_{\rm isotropic}=E(x=0)+\Delta_c \; x+(1/4)W \; x^3+V_{\rm eff}\;  x^2+\ldots
\label{eq:Eisotropic-emery}
\eeq
where $V_{\rm eff} \propto V_{pp}$ is an effective coupling for holes on intersecting rows and columns. 
Clearly, for $x$ small enough $E_{\rm isotropic} > E_{\rm nematic}$. Therefore, at low enough doping the ground state of the Emery model in the strong 
coupling limit is a nematic, in the sense that it breaks spontaneously the point symmetry group of the square lattice, $\mathcal{C}_4$.\footnote{It also turns 
out that in the (so far physically unrealizable) case of $x=1$, the ground state is a nematic insulator as each row is now full. However, for $x \to 1$ the 
ground state is again a nematic metal.} What happens for larger values of $x$ is presently not known. Presumably a complex set of stripe phases exist. 
How this analysis is modified by the spin degrees of freedom is an open important problem which may illuminate our understanding of the physics of high 
temperature superconductors.

The nematic state we have found to be the exact ground state is actually maximally nematic: the nematic order parameter is $1$.  To obtain 
this result we relied on the fact that we have set $t_{pp}=0$ as this is by far the smallest energy scale. 
The state that we have found is reminiscent of the nematic state found in 2D mean field theories of the Pomeranchuk transition\cite{kee-2003} in 
which it is found that the nematic has an open Fermi surface, as we have also found. Presumably, for the more physical case of $t_{pp}\neq 0$, the  strong 
1D-like nematic state we found will show a crossover to a 2D (Ising) Nematic Fermi liquid state.

\section{The Quantum Nematic-Smectic QCP and the Melting of the Stripe Phase}
\label{sec:NSQCP}

We will now turn to the problem of the the quantum phase transition between electron stripe and nematic phases. For simplicity we will consider only the 
simpler case of the charge stripe and the charge nematic, and we will not discuss here the relation with antiferromagnetic stripes and superconductivity. 
Even this simpler problem is not well understood at present.

In classical liquid crystals there are two well established ways to describe this transition, known as the smectic A-nematic transition. One approach is the 
McMillan-de Gennes theory, and it is a generalization of the Landau-Ginzburg theory of phase transitions to this problem (see Ref.\cite{degennes-1993}.) 
The other approach regards this phase transition as a melting of the smectic by proliferation of its topological excitations, 
the dislocations of the smectic order.\cite{nelson-1981,toner-1980,chaikin-1995} 

There are however important (and profound) differences between the problem of the quantum melting of a stripe phase into a quantum nematic and its 
classical smectic/nematic counterpart. The classical problem refers to three-dimensional liquid crystals whereas here we are interested in a two-
dimensional quantum system. One may think that the time coordinate, {\it i.e.\/} the existence of a time evolution dictated by quantum mechanics, provides 
for the third dimension and that two problems may indeed be closely related. Although to a large extent this is correct (this is apparent in the imaginary 
time form of the path integral representation) some important details are different. 

The problem we are interested in involves a metal and hence has 
dynamical fermionic degrees of freedom which do not have a counterpart in its classical cousin. One could develop a theory of quantum melting of the 
stripe phase by a defect  proliferation mechanism by considering only the collective modes of the stripe. 
Theories of this type can be found in Refs.\cite{zaanen-2004,cvetkovic-2006} and in Ref.\cite{radzihovsky-2008} (in the context of a system of cold atoms) 
which lead to several interesting predictions. Theories of this type would describe correctly an insulating stripe state in which the fermionic degrees of 
freedom are gapped and hence not part of the low energy physics. The problem 
of how to develop a non-perturbative theory of this transition with dynamical fermions is an open and unsolved problem.\footnote{Important work with a 
similar approach has been done on the problem of the quantum melting of the stripe state in quantum Hall systems.\cite{wexler-2001,radzihovsky-2002}.}

Another important difference is that in most cases of interest the quantum version of this transition takes place in a lattice system. thus, even if the stripe 
state may be incommensurate, and hence to a first approximation be allowed to ``slide'' over the lattice background, there is no continuous rotational 
invariance but only a point group symmetry leftover. Thus, at least at the lowest energies, the nematic Goldstone mode which plays a key role in the 
classical problem, is gapped and drops out of the critical behavior. However one should not be a purist about this issue as there may be significant 
crossovers that become observable at low frequencies and temperatures if the lattice effects are weak enough. Thus it is meaningful to consider a system 
in which the lattice symmetry breaking are ignored at the beginning and considered afterwards.

In Ref. \cite{sun-2008} a theory of the quantum melting of a charge stripe phase is developed using an analogy with the McMillan-deGennes theory. The 
main (and important) difference is the role that the fermionic degrees of freedom play in the dynamics. Thus we will describe the stripe (which at this level 
is equivalent to a CDW) by its order parameter, the complex scalar field $\Phi(\vec{r},t)$, representing the Fourier component of the charge density operator 
near the ordering wavevector $\vec{Q}$.\footnote{For a different perspective see Ref.\cite{kirkpatrick-2009}.}

We will assume that the phase transition occurs in a regime in which the nematic order is well developed and has a large amplitude $|\mathcal N|$.
 In this regime the fluctuations of the amplitude of the nematic order parameter $N$ are part of the high energy physics, 
 and can be integrated out. In contrast we will assume that the phase mode of the nematic 
order, the Goldstone mode, is either gapless (as in a continuum system) or of low enough energy that it matters to the physics (as if the lattice symmetry 
breaking is sufficiently weak). In this case we will only need to consider the nematic Goldstone (or `pseudo-Goldstone') field which we will denote by 
$\varphi(\vec{r},t)$.

We should note that there is another way to think about this problem, in which one considers the competition  between the CDW order parameters (two in 
this case), the nematic order and the normal Fermi liquid near a suitable multi-critical point. This problem was considered briefly in Ref.\cite{sun-2008} 
and revisited in more detail in Ref.\cite{millis-2009}. The main conclusion is that for a (square) lattice system, the multicritical point is inaccessible as it is 
replaced by a direct (and weak) fluctuation-induced  first-order transition from the FL to the CDW state. Thus, the theory that we discuss here applies far 
from this putative multicritical point in a regime in which, as we stated above, the nematic order is already well developed and large.

Following Ref.\cite{sun-2008} we will think of this quantum phase transition in the spirit of a Hertz-Millis type approach and postulate the existence of an 
order parameter theory coupled to the fermionic degrees of freedom. The quantum mechanical action of the order parameter theory 
$S_{op}[\Phi,\varphi_N]$, 
has the McMillan-deGennes form
\bea
S_{op}&=&|N|^2 \int d\vec{r} dt \Big((\partial_t \varphi_N)^2-K_1 (\partial_x \varphi)^2-K_2(\partial_y \varphi_N)^2\Big)
\nonumber \\
&&\!\!\!\!\!\! \!\!\!\!\!\! \!\!\!\!\!\! +\int d{\vec{r}} dt \left(|\partial_t \Phi|^2-C_y|\partial_y \Phi|^2-C_x \big| \big(\partial_x -i \frac{Q}{2} \varphi_N \big)\big|^2
-\Delta_{CDW} |\Phi|^2-u_{CDW}|\Phi|^4\right)
\nonumber \\
&&
\label{eq:MMdG}
\eea
where $|N|$ is the amplitude of the nematic order parameter, $K_1$ and $K_2$ are the two Franck constants (which were discussed before), 
$C_x$ and $C_y$ are the stiffnesses of the CDW order parameter along the $x$ and $y$ directions, $Q$ is the modulus of the CDW ordering wavevector, 
$\Delta_{CDW}$ and $u_{CDW}$ are parameters of the Landau theory that control the location of the CDW transition ($\Delta_{CDW}=0$) and stability. 
Here we have assumed  a stripe state, a unidirectional CDW, with its ordering wavevector along the $x$ direction. We have also assumed $z=1$ 
(``relativistic'') quantum dynamics which would be natural for an insulating system.

The fermionic contribution has two parts. One part of the fermionic action, $S_{FL}[\psi]$, where $\psi$ is the quasiparticle Fermi field (we are omitting the 
spin indices), describes a conventional Fermi liquid, {\it i.e.\/} the quasiparticle spectrum with a Fermi surface of characteristic Fermi wavevector $k_F$, 
and the quasiparticle interactions given in terms of Landau parameters. What will matter to us is the coupling between the fermionic quasiparticles and the 
nematic order parameter (the complex director field $N$), and the CDW order parameter $\Phi$,
\bea
S_{\rm int}&=&g_N\int d\vec{r} dt \left(\mathcal{Q}_2(\vec{r}, t) N^\dagger(\vec{r}, t)+{\rm h.c.}\right)\nonumber \\
&&+g_{CDW} \int d \vec{r} dt \left(n_{CDW}(\vec{r},t) \Phi^\dagger
(\vec{r},t) +\textrm{h.c.}\right)
\label{eq:S-int}
\eea
where $g_N$ and $g_{CDW}$ are two coupling constants and , as before, 
\beq
\mathcal{Q}_2(\vec{r},t)=\psi^\dagger(\vec{r},t) (\partial_x+i \partial_y)^2 \psi(\vec{r},t)
\label{eq:nematic-fermion}
\eeq
is the nematic order parameter (in terms of quasiparticle Fermi fields), and 
\beq
n_{CDW}(\vec{q},\omega)=\int d \vec{k} d\Omega\;  \psi^\dagger(\vec{k}+ \vec{q}+ \vec{Q}, \omega+\Omega) \psi(\vec{k}, \Omega)
\label{eq:CDW-fermion}
\eeq
is the CDW order parameter (also in terms of the quasiparticle Fermi fields.)

This theory has two phases: a) a nematic phase, for $\Delta_{CDW}>0$, where $\rangle \Phi \langle=0$, and b) a CDW phase, for $\Delta_{CDW}<0$, 
where $\langle \Phi \rangle\neq =0$, separated by a QCP at $\Delta_{CDW}=0$. In the nematic (``normal'') phase the only low energy degrees of freedom 
are the (overdamped) fluctuations of the nematic Goldstone mode $\varphi_N$, and nematic susceptibility in the absence of lattice effects (which render it 
gapped otherwise)
\beq
\chi_\perp^N(\vec{q},\omega)=\frac{1}{g_N^2 N(0)} \frac{1}{\big(i\frac{\omega}{q}\sin^2(2\phi_q)-K_1q_x^2-K_2q_y^2\big)}
\eeq
where $\sin(2\phi_q)=2q_x q_y/q^2$ and $N(0)$ is the quasiparticle density of states at the FS.

We will consider here the simpler case in which the CDW ordering wavevector obeys $Q<2k_F$ (see the general discussion in Ref.\cite{sun-2008}.) In 
this case one can see that the main effect of the coupling to the quasiparticles (aside from some finite renormalizations of parameters) is to change the 
dynamics of the CDW order parameter due to the effects of Landau damping. The total effective action in this case becomes
\bea
S&=&\int \frac{d \vec{q} d \omega}{(2\pi)^3}\; C_0 i |\omega| |\Phi(\vec{q},\omega)|^2\nonumber \\
&&\!\!\!\!\!\! \!\!\!\!\!\! -\int d \vec{r} dt\; \left(C_y |\partial_y \Phi |^2+C_x \big|\left(\partial_x -i \frac{Q}{2} \varphi_N\right)\Phi|^2+\Delta_{CDW} |\Phi|^2+u_{CDW}|\Phi|^4\right)
\nonumber \\
&&\!\!\!\!\!\! +\int \frac{d \vec{q} d\omega}{(2\pi)^3} \; \left(\tilde K_0 \frac{i |\omega|}{q} \sin^2(2\phi_q)-\tilde K_1 q_x^2-\tilde K_2 q_y^2\right)
|\varphi_N(\vec{q},\omega)|^2
\nonumber \\
&&
\label{eq:S-N-CDW}
\eea
where $C_0 \sim g_{CDW}^2$, $\tilde K_0= g_N^2 |N|^2 N(0)$ and $K_{1,2}=g_N^2|N|^2 N(0) K_{1,2}$.

By inspecting Eq.\eqref{eq:S-N-CDW} one sees that at $\Delta_{CDW}=0$, as before the nematic Goldstone fluctuations have $z=3$ 
(provided they remain gapless), and the CDW 
fluctuations have $z=2$. Thus the nematic Goldstone modes dominate the dynamics at the nematic-CDW QCP. Even if 
the nematic Goldstone modes were to become gapped (by the lattice anisotropy), the QCP now will have $z=2$ (due to Landau damping) instead of 
$z=1$ as in the ``pure'' order parameter theory. In both cases, the nematic Goldstone mode and the CDW order parameter fluctuations effectively 
decouple in the nematic phase. The result is that the nematic phase has relatively low energy CDW fluctuations with a dynamical susceptibility
\beq
\chi^{CDW}(\vec{q},\omega)=-i \langle \Phi^\dagger(\vec{q},\omega)\Phi(\vec{q}, \omega)\rangle_{\rm ret}=
\frac{1}{i C_0  |\omega|-C_x q_x^2- C_y q_y^2-\Delta_{CDW}}
\label{eq:fluctuating-cdw}
\eeq
In other terms, as the QCP is approached, the nematic phase exhibits low energy CDW fluctuations that would show up in low energy inelastic scattering 
experiments much in the same was as the observed {\em fluctuating stripes} do in inelastic neutron scattering experiments in the high temperature 
superconductors.\cite{kivelson-2003} As we saw before, a regime with ``fluctuating'' CDW (stripe) order is a nematic.

A simple scaling analysis of the effective action of Eq.\eqref{eq:S-N-CDW} shows that, since $z>1$ at this QCP, the coupling between the nematic 
Goldstone mode $\varphi_N$ and the CDW order parameter $\Phi$ is actually {\em irrelevant}. In contrast, in the (classical) case it is a marginally relevant 
perturbation leading to a fluctuation induced first order transition.\cite{halperin-1974,chaikin-1995} Thus, this ``generalized McMillan-de Gennes'' theory 
has a continuous (quantum) phase transition which, possibly, may become weakly first order at finite temperature. 

This is not to say, however, that the stripe-nematic quantum phase transition is necessarily continuous. In Ref.\cite{sun-2008} it is shown that the nature of 
the quantum phase transition depends on the relation between the ordering wave vector $\vec{Q}$ and the Fermi wave vector $k_F$. For $|\vec{Q}|<2k_F$ 
the transition is continuous and has dynamical scaling $z=2$. Instead, for $|\vec{Q}|=2k_F$ it depends on whether $|\vec{Q}|$ is commensurate or 
incommensurate with the underlying lattice: for the incommensurate case the transition is (fluctuation induced) first order (consistent with the results of 
Ref.\cite{altshuler-1995}) but it is continuous for the commensurate case with $z=2$ and anisotropic scaling in space.

As in the case of the Pomeranchuk transition, the quasiparticles are effectively destroyed at the stripe-nematic QCP as well. indeed, already to order one 
loop it is found\cite{sun-2008} that the quasiparticle scattering rate scales with frequency as $\Sigma^{''}(\omega) \propto \sqrt{|\omega|}$, signaling a 
breakdown of Fermi liquid theory. As in our discussion of the nematic-FL QCP, this behavior must be taken as an indication of a breakdown of 
perturbation theory and not as the putative ultimate quantum critical behavior, which remains to be understood.

In the quasiparticle picture we are using, the stripe state is similar to a CDW. Indeed, in the broken symmetry state the Fermi surface of the nematic is 
reconstructed leading to the formation of fermion pockets. As we noted above, we have not however assumed a rigid connection between the the ordering 
wave vector and the Fermi surface and, in this sense, this si not a weak coupling state. Aside from that, in the presence of lattice pinning of the 
nematic Goldstone mode, the asymptotic lowe energy properties of the stripe state are similar to those of a CDW (for details, see Ref.\cite{sun-2008}.)

\section{Outlook}
\label{sec:outlook}

In these lectures we have covered a wide range of material on the theory of electronic liquid crystal phases and on the experimental evidence for them. As 
it is clear these lectures have a particular viewpoint, developed during the past decade in close collaboration with Steven 
Kivelson. I have tried, primarily at the level of citations as well an on numerous caveats, to make it clear that there are 
many important unsolved and still poorly understood questions that (at present) allow for more than one answer. It is a problem that requires the 
development of many points of view which eventually complement each other. 

Several major problems remain open. One of them, in my view the most pressing one, is to establish 
the relation between the existence of these phases (stripes, nematics, 
etc.) and the mechanism(s) of high temperature superconductivity. In my opinion there is mounting experimental evidence, some of which I discussed
 here, that strongly suggests the existence of a close and probably
unavoidable connection. A question that deserves more consideration is the particular connection between nematic order and superconductivity. 
Superficially these two issues would seem quite orthogonal to each other. Indeed, it is hard to see any connection within the context of a weak coupling 
theory. However if the nematic order arises from melting a stripe state which has a spin gap (such as the pair density wave state we discussed in these 
lectures) it is quite likely that a close connection may actually exist and be related. The current experimental evidence suggests such a relation.

Another key theoretical question that is wide open is to develop a more microscopic theory of the pair density wave state. In spite with the formal analogy 
with the Larkin-Ovchinnikov state, it seems very unlikely that a a ``straight BCS approach'' would be successful in this case. This state seems to have a 
strong coupling character.

As it must be apparent from the presentation of these lectures, the theory that has been done (and that is being done now) is for the most part quite 
phenomenological in character. There are very few ``rigorous'' results on the existence of these phases in  strongly correlated systems. The notable 
exception are the arguments we presented for the existence of nematic order in the strong coupling limit of the Emery model. Clearly more work is 
needed.

\textit{Acknowledgments}: I am deeply indebted to Steve Kivelson with whom we developed many of the ideas that are presented here.  
Many of these results  were obtained also in collaboration with my former students Michael Lawler and Kai Sun, as well as to John Tranquada, 
Vadim Oganesyan, Erez Berg, Daniel Barci, Congjun Wu, Benjam{\'\i}n Fregoso, Siddhartha Lal and Akbar Jaefari, and many other collaborators. 
I would like to thank Daniel Cabra, Andreas Honecker and Pierre Pujol for  
inviting me to this very stimulating Les Houches Summer School on ``Modern theories of correlated electron systems'' (Les Houches, May 2009). 
This work was supported in part by the National Science Foundation, under grant DMR 0758462 at the University of Illinois, 
and by the Office of Science, U.S. Department of Energy, under Contract DE-FG02-91ER45439 through the Frederick
Seitz Materials Research Laboratory of the University of Illinois.

\vspace{1cm}


\end{document}